\documentclass[fleqn,usenatbib]{mnras}
\defcitealias{hutschenreuter22}{H22}
\defcitealias{unger24}{UF24}

\usepackage{newtxtext,newtxmath}

\usepackage[T1]{fontenc}

\DeclareRobustCommand{\VAN}[3]{#2}
\let\VANthebibliography\thebibliography
\def\thebibliography{\DeclareRobustCommand{\VAN}[3]{##3}\VANthebibliography}

\usepackage{graphicx}	
\usepackage{amsmath}	

\title[The radial component of the local GMF]{The radial component of the local Galactic magnetic field in 3D}

\author[L. McCallum et al.]{
Lewis McCallum,$^{1}$\thanks{E-mail: mccallum@physik.rwth-aachen.de}
Philipp Frank,$^{2}$
Sebastian Hutschenreuter,$^{3}$
Robert Benjamin,$^{4}$
Rebecca A. Booth,$^{5}$
\newauthor
Susan E. Clark,$^{6,2}$
Marijke Haverkorn,$^{7}$
Alex S. Hill,$^{8,9}$
Philipp Mertsch,$^{1}$
Anna Ordog,$^{8,9,10}$
\newauthor
Andrew K. Saydjari,$^{11}$\thanks{Hubble Fellow}
Jennifer West$^{9}$
\\
$^{1}$Institute for Theoretical Particle Physics and Cosmology, RWTH Aachen University, Sommerfeldstraße 16, 52074 Aachen, Germany\\
$^{2}$Kavli Institute for Particle Astrophysics \& Cosmology, P.O. Box 2450, Stanford University, Stanford, CA 94305, USA\\
$^{3}$University of Vienna, Department of Astrophysics, T\"urkenschanzstra{\ss}e 17, 1180 Vienna, Austria \\
$^{4}$University of Wisconsin, Whitewater, Whitewater, WI, USA \\
$^{5}$Department of Physics and Astronomy, University of Calgary, Calgary, Alberta, Canada \\
$^{6}$Department of Physics, Stanford University, Stanford, CA 94305, USA \\
$^{7}$Department of Astrophysics/IMAPP, Radboud University, PO Box 9010, 6500 GL Nijmegen, The Netherlands \\
$^{8}$Department of Computer Science, Math, Physics, \& Statistics, University of British Columbia, Okanagan Campus, Kelowna, BC, Canada \\
$^{9}$Dominion Radio Astrophysical Observatory, Herzberg Research Centre for Astronomy and Astrophysics, National Research Council, Penticton, BC, Canada \\
$^{10}$Department of Physics \& Astronomy, University of Western Ontario, 1151 Richmond Street, London, ON, Canada \\
$^{11}$Department of Astrophysical Sciences, Princeton University, Princeton, NJ 08544, USA \\
}

\date{Accepted XXX. Received YYY; in original form ZZZ}

\pubyear{\the\year{}}

\begin{document}
\label{firstpage}
\pagerange{\pageref{firstpage}--\pageref{lastpage}}
\maketitle

\begin{abstract}
We present a distance-resolved reconstruction of the local line-of-sight Galactic magnetic field, $B_{||}$, by combining a 3D electron density ($n_{e}$) map derived from dust map-informed simulations and a full-sky map of Faraday rotation measure (RM). The forward model evaluates RM on the same 3D grid as the $n_{e}$ map and compares to the Galactic Faraday rotation sky. We infer $B_{||}$ with a Gaussian-process prior whose power spectrum is inferred from the data using geometric variational inference. The result is a local (within 1.25 kpc where $|b|>5^{\circ}$) map of $B_{||}$ with uncertainties. The reconstructed RM sky reproduces prominent features of Faraday rotation sky, with a root mean square average strength of $B_{||}$ of $1.63\pm 0.16\rm \mu G$. In face-on views, the magnetic field exhibits coherent patches with alternating sign and hints of kpc-scale modulations, but with significant structure seen on scales of order 100~pc. The $B_{||}$ field is seen to exhibit a 3D power spectrum with an average slope of $-2.73 \pm 0.19$. We validate our $B_{||}$ reconstruction with Galactic pulsars. Predicted RMs (computed by integrating $n_{e}B_{||}$ to each pulsar’s distance) correlates with observed RMs, and predicted dispersion measures (DMs) from the $n_{e}$ map also correlate with measured DMs, albeit with significant scatter. 
\end{abstract}




\section{Introduction}

Magnetic fields are known to pervade the Milky Way, and serve many roles in shaping the physics of the interstellar medium (ISM) \citep{beck15}. They provide pressure support against the gravitational potential \citep{boulares90}, affect the collapse of molecular clouds and star-formation \citep{crutcher12,klos25}, guide cosmic ray transport, and couple to kinetic turbulence across different scales and phases of the ISM \citep{elmegreen04}. Observationally, the Galactic magnetic field (GMF) is often considered to be comprised of a large scale, coherent structure plus a random, turbulent component, with total magnetic field strengths typically a few to tens of $\mu G$ \citep{haverkorn15}. The GMF can be observed through a variety of tracers, including synchrotron polarization, dust polarization, Zeeman splitting and Faraday rotation \citep{planck16,jaffe19}, the latter of which is the focus of this work.

A large body of work has mapped the Galaxy’s magnetic field with global, parametric models constrained by extragalactic RMs and synchrotron/dust emission \citep{page07,sun08,sun10,jaffe10,pshirkov11,fauvet12,jansson12,terral17,han18,unger24,korochkin25}, delivering kpc-scale trends but relying on analytic assumptions about disk/halo geometry, spiral arms, symmetry, individual objects and sometimes turbulent components. Tests from \citet{planck16} showed that fitting both synchrotron and dust data with a single global model is challenging, with results often sensitive to assumptions about cosmic ray electrons and component separation in the Planck microwave bands. \citet{hutschenreuter24} provides a non-parametric model for the magnetic field using tracers of the Galactic electron density. However, these efforts are confined to a projection of the magnetic field on the sky and not in 3D.  Other efforts have been made in the 3D mapping of the plane-of-sky component of the Galactic magnetic field using starlight polarization and galactic synchrotron emission \citep{pelgrims24}.

Despite these advances, a distance-resolved, non-parametric 3D reconstruction of the local GMF from RM data remains largely unexplored. In this work, we combine the 3D dust map-based grids of $n_{e}$ \citep{edenhofer23,mccallum25} with the RM sky \citepalias{hutschenreuter22} in an Information Field Theory (IFT) \citep{enslin09} framework to recover a distance-resolved map of the line of sight component of the magnetic field ($B_{||}$) in the local ($d < 1.25$ kpc) volume at <10-pc resolution.

Our reconstruction relies on the previously reconstructed Faraday RM sky of \citet{hutschenreuter22} (herein referred to as \citetalias{hutschenreuter22}). The Faraday effect is the frequency dependent rotation of the polarization plane of linearly polarized light in the presence of the magnetized plasma. For the case of a a background source, with rotation happening in the intervening medium between the source and observer (sometimes referred to as a `Faraday screen'), the RM can be considered as described in \citet{ferriere21}:

\begin{equation}
\mathrm{RM} \;=\; \frac{e^{3}}{2\pi m_{e}^{2}c^{4}} \int_0^{d} n_e (s)\,B_{\parallel}(s)\,\mathrm{ds},
\label{eq:rm_forward}
\end{equation}

where $s$ is the coordinate along the line of sight, measured from the source at 
$s=0$ to the observer at $s=d$. The quantity $n_{e}(s)$ is the free electron number density and $B_{||}(s)$ is the component of the magnetic field along the line of sight, both of which vary with position. We adopt the convention that $B_{||} > 0$ when the magnetic field component points toward the observer. The RM can thus be simply described as the path integral of $0.81n_{e}(s)B_{||}(s)$, and it is linearly sensitive to both the electron density and the line-of-sight magnetic field. While this measurement gives insight into both components, this also introduces a degeneracy between contributions from $B_{||}$ versus $n_{e}$. In this work, we address this degeneracy by using the recent 3D maps of electron density from \citet{mccallum25}. We also exploit the full-sky reconstruction of the Galactic Faraday rotation sky from \citetalias{hutschenreuter22}, which assimilates essentially all extragalactic RM measurements available by the end of 2020 \citep{van_eck23} into a homogeneous map with uncertainties.

We carry out the inference of $B_{||}$ as a Bayesian field reconstruction problem within IFT and solve it with the python package \texttt{NIFTY8} \citep{selig13,steininger19,edenhofer24_nifty}, using variational inference (VI) to approximate the posterior over high-dimensional fields. Such IFT/VI approaches have recently powered several high-dimensional reconstructions (such as \citetalias{hutschenreuter22}), as well as specifically 3D ISM reconstructions \citep{leike20,edenhofer23,soding25}. We utilize closely related techniques here.

Our method yields a distance-resolved map of $B_{||}$ with quantified uncertainties. We validate our inferred 3D structure of RM using Galactic pulsars with measured RM and parallax based distances. Along pulsar sight-lines, we integrate the reconstructed $B_{||}n_{e}$ to each pulsar distance, providing an independent, physically interpretable test of the model. We also validate our method by carrying out a mock reconstruction using a synthetic map of $B_{||}$, and carrying out further tests of the reliability of small scale and radially resolved structures.

\begin{figure*}
    \centering
    \includegraphics[width=1.0\textwidth]{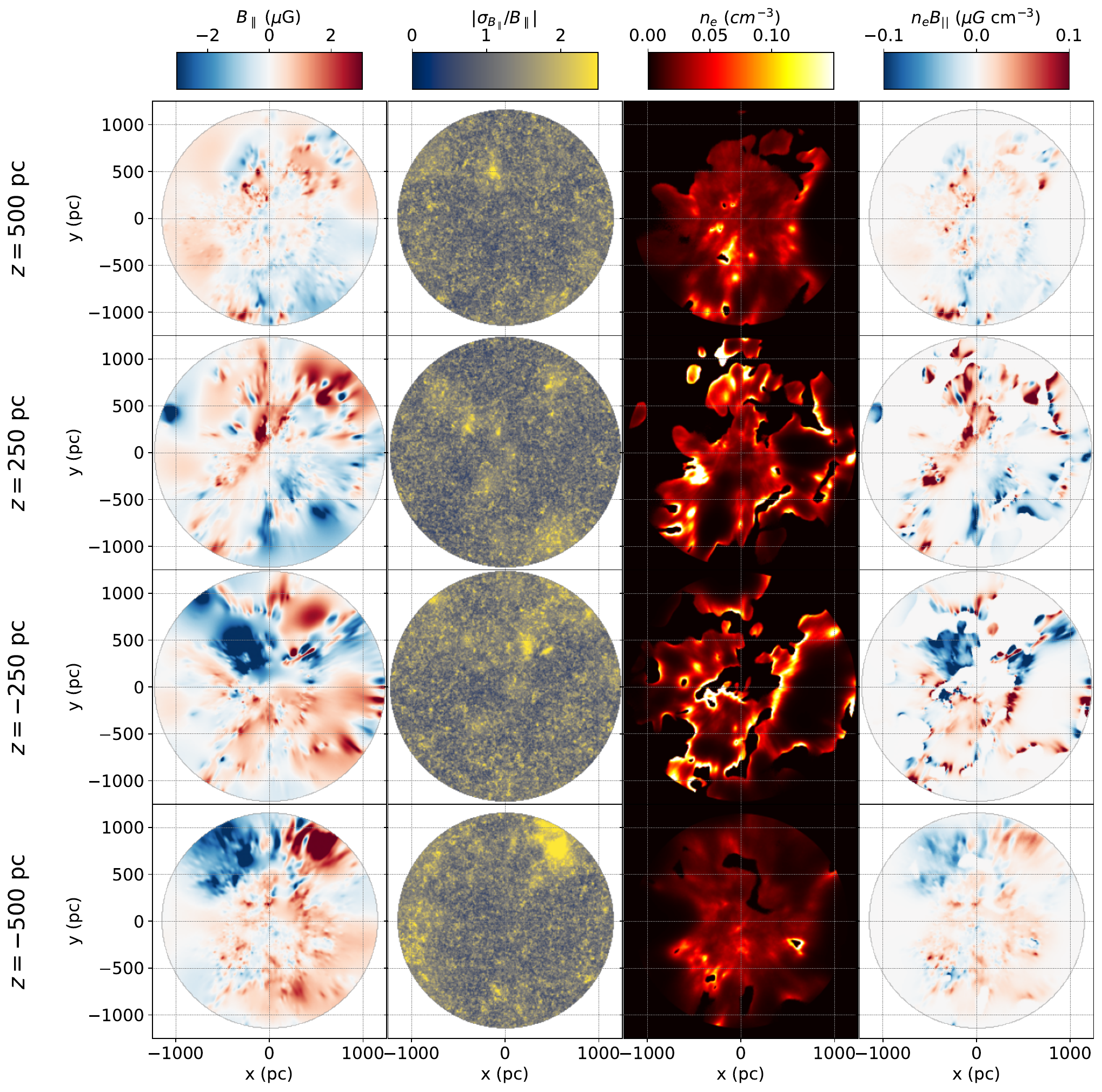}
    \caption{Selected slices through our reconstructed fields. These slices are one grid cell thick (approximately 10~pc). The left column shows the inferred structure of $B_{||}$, the second column shows slices of the relative standard deviation of $B_{||}$ as determined from our 120 samples, third column shows the electron density structure used for our reconstruction, and the rightmost column shows the 3D structure of RM contributions. All slices are in the $x-y$ plane, at Z-heights of $500$~pc, $250$~pc, $-250$~pc and $-500$~pc. Note that the $B_{||}$ field is only showing the strength of $B_{||}$ relative to the position of the sun, and that large values of $B_{||}$ at high altitude have a significant component point in the $z$-axis as well as in the $x-y$ plane. }
    \label{fig:allslices}
\end{figure*}

A high-resolution 3D map of the Galactic magnetic field would be broadly enabling. It would allow for more precise back-tracking and deflection modelling of ultra-high energy cosmic rays (UHECRs); constrain Galactic cosmic-ray transport; provide line-of-sight aware predictions for synchrotron emission; and strengthen Cosmic Microwave Background (CMB) foreground separation by anchoring synchrotron and dust polarization in real space. Beyond these, it could subtract large-scale Milky Way contributions to pulsar/fast radio burst (FRB) RM–DM studies \citep{pandhi22}, reduce foreground systematics for 21-cm Epoch of Reionization studies, and supply a ground truth for magnetised ISM models of star formation, feedback, and disk–halo coupling. 

Such a map would also enable magnetic structure to be tracked continuously across physical scales, from parsec-scale turbulence to kiloparsec-scale Galactic organization, helping to bridge the multi-scale picture inferred from Faraday-rotation. At the same time, a distance-resolved reconstruction could directly constrain the locations of major magnetised radio and RM features whose distances remain uncertain, such as large spurs, shells, and local-bubble-scale structures.

In section~\ref{methods} we describe the utilised data and methods, section~\ref{results} shows our results, section~\ref{discussion} discusses the results, comparing to other models and analysing the reliability of our reconstruction, and section~\ref{conclusions} shows our conclusions.


\begin{figure*}
    \centering
    \includegraphics[width=1.0\textwidth]{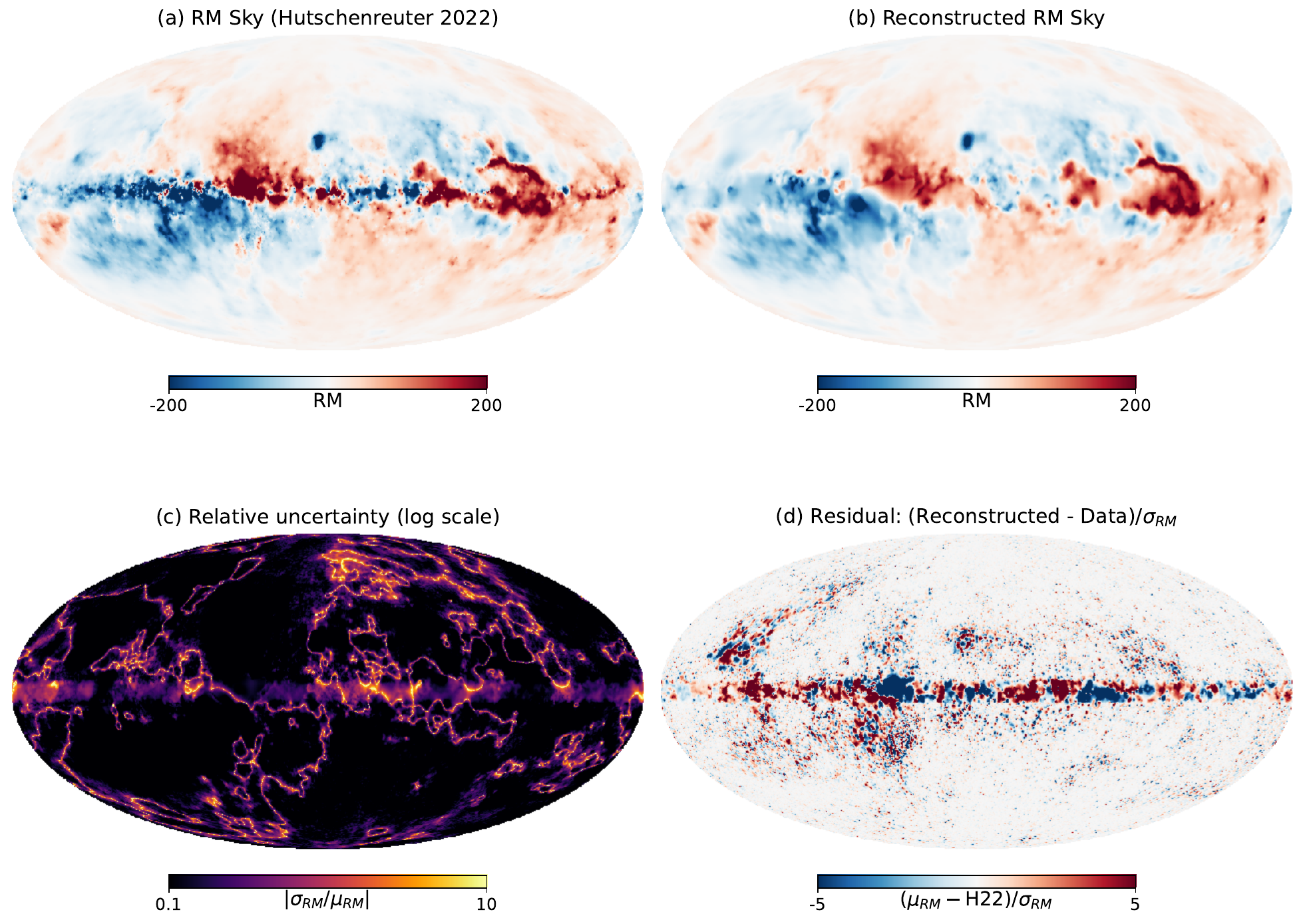}
    \caption{Maps derived from our reconstructed magnetic field structure in Galactic coordinates centred at $(l,b) = (0^{\circ},0^{\circ})$. Panel (a): the data-constrained extragalactic RM sky from \citetalias{hutschenreuter22}. Panel (b): our reconstruction of the RM sky from our inferred 3D structure of $B_{||}$ and underlying 3D map of $n_{e}$. This map is derived from the mean $B_{||}$ field from the posterior samples. Panel (c): the total uncertainty in the recovered map of RM, defined as the pixel-by-pixel standard deviation from each of the posterior samples of our reconstruction. This takes into account the underlying uncertainties in the \citetalias{hutschenreuter22} map, uncertainties in the form of samples from the \citet{edenhofer23} dust map, and sample variance originating from the differing draws from the posterior. Panel (d): the normalized residual sky as a difference in RM in our posterior mean sky map and the sky of \citetalias{hutschenreuter22}, normalized to the uncertainty in the reconstructed sky.}
    \label{fig:reconstructed}
\end{figure*}

\section{Methods}
\label{methods}

\begin{figure*}
    \centering
    \includegraphics[width=1.0\textwidth]{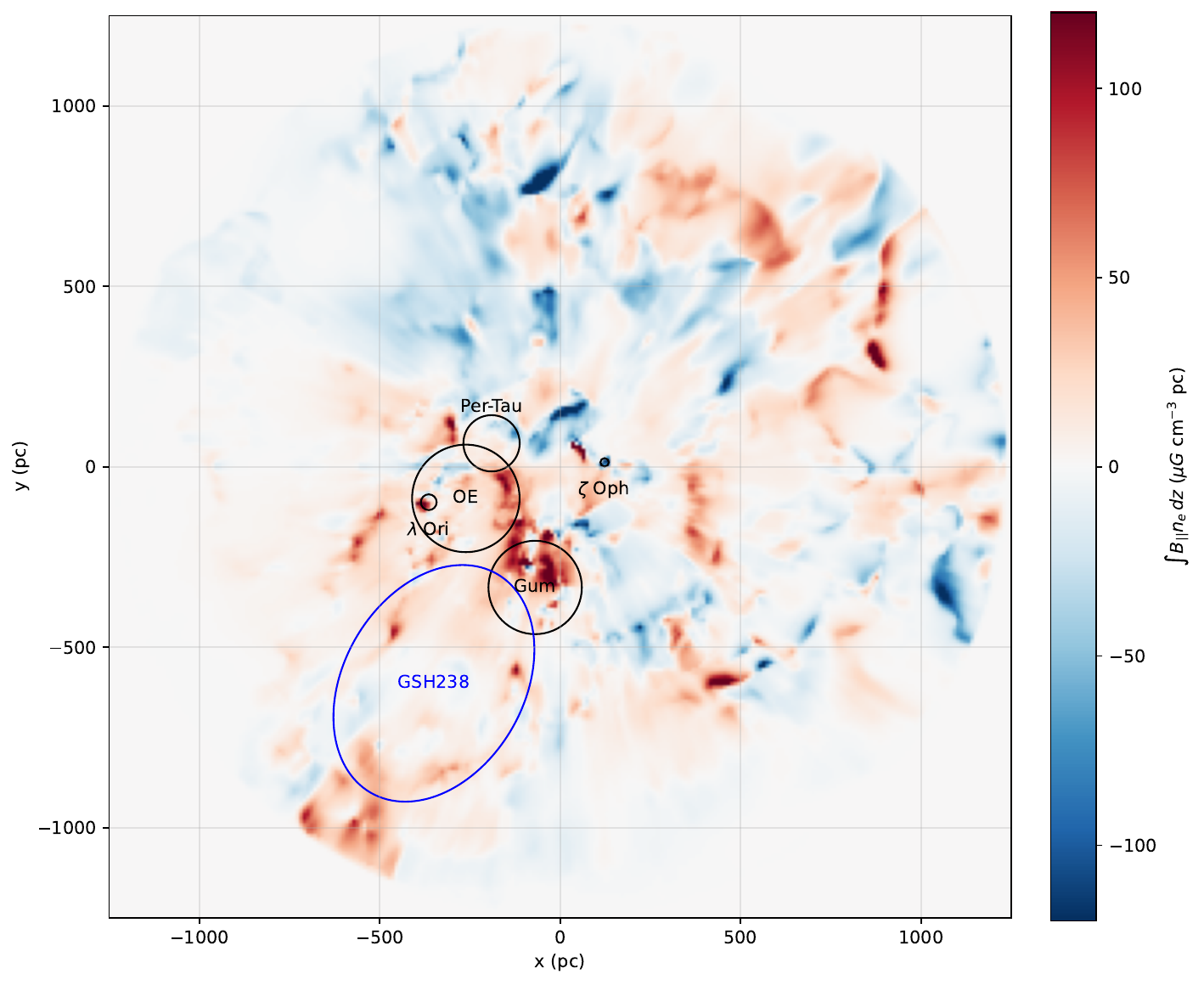}
    \caption{The mean reconstructed 3D field of $n_{e} B_{||}$ integrated along the full $z$-axis of the reconstructed volume (excluding the $|b|<5^{\circ}$ masked mid-plane). This corresponds to a z-integrated contribution to the RM sky, shown in a face-on (top-down) view of the local volume. The Sun is at the centre of this image, and the Galactic centre is defined as being to the right. Positive $B_{||}$ is defined as pointing towards the Sun. Any voxels of the 3D grid which are behind our $|b| < 5^{\circ}$ midplane mask are not included in this integration. This can be considered a top down view of the Sun-centred \citetalias{hutschenreuter22} RM map, minus the midplane behind the $5^{\circ}$ mask. Some individual objects are labelled as in \citet{mccallum25}, with the size and location of the Per-Tau shell from \citet{bialy21}. Note that this is not equivalent to the RM which would be observed through the Galactic disc from above, as the direction of the parallel component of the B-field is here still defined using a heliocentric observer. An animated version of this figure is available online as supplementary material, showing the posterior mean from each electron density sample.}
    \label{fig:meanbfield}
\end{figure*}

\subsection{Data}

For a known 3D structure of $n_e$, and a 3D description of where Faraday rotation is occurring, we can trivially infer the structure of $B_{||}$. Throughout this paper, we will often refer to Faraday rotation in terms of its local contribution per unit path length,
$d{\rm RM}/ds \propto n_e B_{||}$, so that the ``3D Faraday rotation field'' in what follows refers to this local quantity (which is proportional to the field $n_e B_{||}$), whose integral along a line of sight yields the observed RM.

Recent advances in the 3D mapping of the Galaxy in dust extinction \citep{edenhofer23} have enabled static photoionization simulations which approximate the 3D structure of free electrons within 1.25~kpc of the Sun \citep{mccallum25}. Unfortunately, reconstructing the 3D structure of RM at parsec scale resolution from direct pulsar measurements is infeasible due to a dearth of data (fewer than 100 individual pulsar RM measurements within 1~kpc). Because of this, we seek to model the 3D contributions of $B_{||}$ using the full-sky map of RM from \citetalias{hutschenreuter22}, and distance information from the highly resolved 3D structure of $n_{e}$, which in turn uses tens of millions of \emph{Gaia} extinction measurements. Below, we outline our utilized data, forward model and methods of inference.
\subsubsection{3D Electron Density Structure}
\label{data:electron}

For the 3D structure of free electron density, we use the map from \citet{mccallum25}. These maps are derived from the 3D differential dust extinction maps of \citet{edenhofer23}, combined with the positions and ionizing luminosities of the 87 known O stars and 1 Wolf-Rayet star within the reconstructed volume. The dust map is interpolated onto a Cartesian grid of $1024\times 1024\times1024$ gridcells, spanning 1.25~kpc in each direction for a total box length of 2.5~kpc in each axis. While the reconstructed volume is that of a cube of side length 2.5~kpc, the \citet{edenhofer23} dust map only reaches a distance of 1.25~kpc in each direction, and hence our $n_{e}$ map is only valid in a the central 1.25~kpc radius sphere of the volume. There is also no dust data within a radius of 69~pc, so we also do not have $n_{e}$ structure within this smaller inner sphere.

The inner 69~pc radius sphere does not contribute to our reconstruction, and is considered `empty'. This is a limitation of the dust map and thus the electron density map. While this is a clear source of systematics, the inner volume of the local bubble is of relatively low density, so will contribute significantly less to the RM sky than other regions of our reconstruction. This low density region accounts for only 5\% of each LOS and 0.017\% of the total reconstructed volume.

\citet{mccallum25} assumes a fixed dust to gas ratio in order to map dust extinction to hydrogen number density, and a fixed helium abundance of 0.1. The population of 88 ionizing stars within the reconstructed volume is used as an ionizing photon source distribution, which is traversed in the form of Monte Carlo photon packets through the grid of hydrogen and helium density using the Monte Carlo Radiative Transfer (MCRT) code, \texttt{CMacIonize} \citep{cmi1,cmi2}.

These MCRT packets are used to calculate the equilibrium ionization state, and equilibrium temperature of the O star photoionized gas. With the ionization state of the photoionized gas known, the density of free electrons can be calculated as the sum of ionized hydrogen density and singly ionized helium density. Doubly ionized helium does not contribute to these electron counts, as its ionization potential is higher than the most energetic photons traced by the \texttt{CMacIonize} code. These 3D maps of electron density can be converted to 2D synthetic sky maps of H$\alpha$, and show a good morphological match to the observed H$\alpha$ sky.

While the \citet{mccallum25} electron density map is published with a resolution of $1024^{3}$ gridcells over the 2.5~kpc box, in order to fit our inference onto the GPU hardware available, we degrade the resolution of the map to $256^{3}$.

While the electron density map of \citet{mccallum25} is the most resolved 3D map of $n_{e}$ available, allowing for our non-parametric reconstruction of the magnetic field, this data product is shipped with a number of known caveats and systematics. The most relevant of these are 1) the assumption of a constant gas-to-dust ratio in determining the gas density from the dust extinction map, and 2) missing sources of ionization such as B stars, sdOBs, hot white dwarfs, and cosmic-rays. The constant gas-to-dust assumption will in some places bias the electron density high, and in others low, while the missing ionization sources result in some regions of zero electron density. Where the electron density is systematically overestimated, our magnetic field reconstruction will show lower field strengths to compensate. The opposite is true in regions of low (or zero) modelled electron density. \citet{mccallum25} also notes a strong sensitivity of the modelled $n_{e}$ map on the positions of ionizing sources. If an O-star is placed outside the dense cloud in which it truly resides, it can erroneously ionize a vastly different volume than if it were contained within the dense medium. This effect proves difficult to fold in to the magnetic field reconstruction of this work, and is left as another known source of systematic uncertainty in the underlying $n_{e}$ maps.

\begin{figure}
    \centering
    \includegraphics[width=1.0\columnwidth]{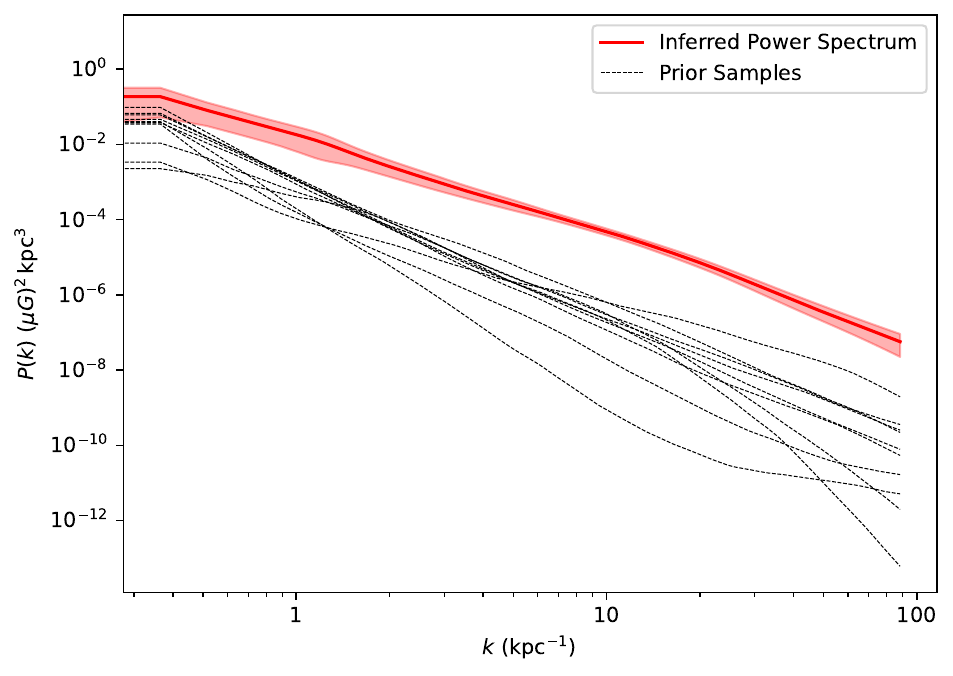}
    \caption{The mean power spectrum of $B_{||}$ from our posterior samples, with the filled region representing $1\sigma$. The mean average slope of the power spectrum is $-2.73 \pm 0.19$. Also shown in black dashed lines are 10 samples from our prior.}
    \label{fig:powerspec}
\end{figure}

\begin{figure*}
    \centering
    \includegraphics[width=1.0\textwidth]{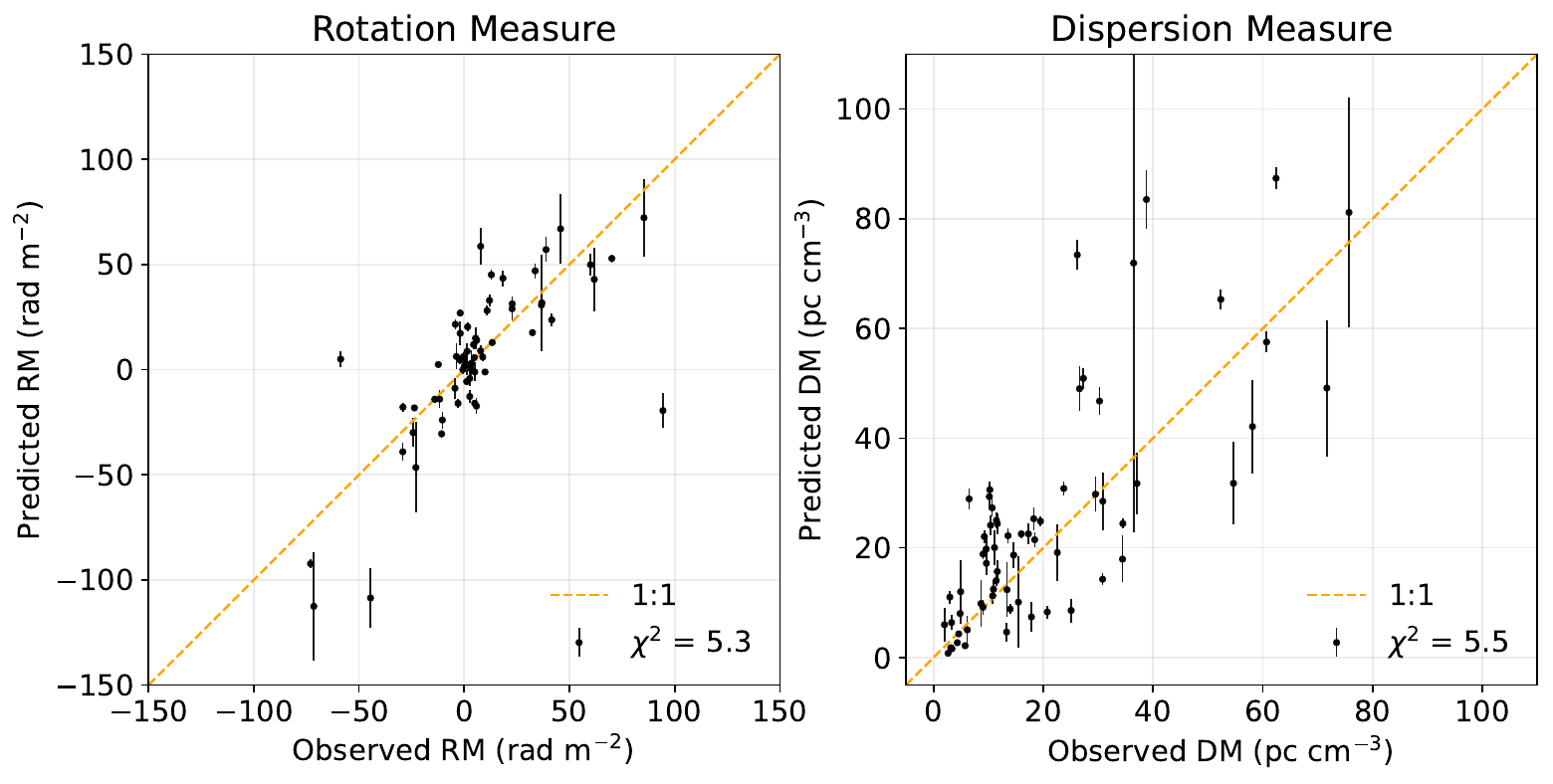}
    \caption{Validation results using local pulsars from the ATNF Pulsar Catalogue \citep{manchester05}. Left panel shows the observed RM of each pulsar versus the RM predicted from the 3D GMF $B_{||}$ reconstruction. The right panel shows the observed DM (scales only with $n_{e}$) with the prediction from the underlying electron density map used as a fixed prior in our reconstruction. 1:1 match lines are shown on each panel. Observational uncertainties in RM and DM are negligible on the scales of these plots, so these error bars are not included.}
    \label{fig:rmdm}
\end{figure*}

\subsubsection{The Faraday Rotation Measure Sky}

For the RM constraint we adopt the all-sky reconstruction of the Galactic Faraday rotation sky by \citetalias{hutschenreuter22}, along with their reported pixel-by-pixel uncertainties. This work compiles 55,190 extragalactic RMs available by the end of 2020, and reconstructs the Galactic RM using IFT with an uncertainty-aware likelihood and correlation model. The resulting product is a homogenized full-sky RM map, alongside a corresponding map of uncertainties.

The \citetalias{hutschenreuter22} reconstruction algorithm contains a statistical noise estimation procedure that takes into account systematic biases stemming from extragalactic, ionospheric or instrumental contaminations.
Since the underlying RM catalog is not homogenously distributed on the sky, the effective angular resolution of this sky map varies, with especially the celestial southern sky being undersampled. This effect will imprint itself on our reconstructions of the magnetic field. In order to fit our reconstruction within our available GPU's memory limits, we degrade the RM map to a HEALPix \citep{gorksi05} grid of $N_{side}=256$. 

While our electron density structure captures only the contributions from within 1.25~kpc, the used RM sky includes Galactic contributions from much further than this. In order to eliminate regions where the RM might be dominated by non-local $n_{e}B_{||}$, we apply a mask to the data at latitudes of $|b| < 5^{\circ}$, and this region is not used in the fitting. This is equivalent to the exclusion of regions $|z| < 100~\rm pc$ at a distance of 1.25~kpc. While the correlation structure of the Gaussian random field may be able to inform structures in this volume without data constraints, we consider the volume inside this hidden segment of the sky to be unconstrained in $B_{||}$, and in this work will not draw conclusions from reconstructed fields in this region. The latitude limit of $5^{\circ}$ was determined from experimentation with this parameter, and this was found to be the lowest value which masks out regions that consistently produce significant variance in the reconstructed sky posterior samples.

While contamination from structures outside our $d<1.25~\rm kpc$ electron map is a likely source of systematics, simulations of Milky Way--like disc galaxies by \citet{pakmor18} found that the RM sky as observed from the solar circle is dominated by the local environment, and largely by the Local Bubble \citep{Maconi_2025}. This result is also observationally corroborated by \citet{korochkin25} and \citet{pelgrims25}, justifying our use of the \citetalias{hutschenreuter22} sky in combination with the local ($<1.25~\rm kpc$) electron density map.

\subsubsection{Individual RM Measurements - Local Galactic Pulsars}

Galactic pulsars with known parallaxes provide an independent, distance-constrained probe of the line-of-sight magnetic field. For any 3D reconstruction of $B_{||}$, we can compute the model-predicted RM for each pulsar and compare to the observed RM. Because these objects were not used in the fit, agreement between these values constitutes a test of accuracy and predictive power.

As well as RMs, many of the locally observed pulsars have observed dispersion measures (DMs). Pulsar DMs are a direct measure of the intervening electron column density between the source and observer. DM is defined as:
\begin{equation}
    DM = \int_{0}^{d} n_{e}(s) ds
\end{equation}

again where $s$ is the coordinate along the line of sight, measured from the source at $s=0$ to the observer at $s=d$.

Just as the individual RMs give us a test of our reconstructed field of $n_{e}B_{||}$, the observed DMs give us a test of underlying assumed $n_{e}$ maps.

We evaluate the predicted pulsar DMs by integrating the electron density map, and RMs by integrating our reconstructed $B_{||}n_{e}$. Our modelled DMs carry an uncertainty which is based on original uncertainties from the \citet{mccallum25} electron density map and the uncertainties in the pulsar distances. The modelled RMs contain uncertainties obtained from the variance in our $B_{||}$ reconstruction samples (which carries forward the uncertainties in the RM sky map from \citetalias{hutschenreuter22}), the uncertainties in the underlying electron density map, and also the distance uncertainties to the pulsars. Electron density map uncertainties are based on the uncertainty in the \citet{edenhofer23} dust map.

Pulsar positions, RMs, DMs, and available distance information in parallaxes were taken from the ATNF Pulsar Catalogue \citep{manchester05}\footnote{\url{http://www.atnf.csiro.au/research/pulsar/psrcat}} for all pulsars with parallaxes placing them within 1.25~kpc.

\subsection{Inference of $B_{\parallel}$}

Our goal is to infer the line-of-sight magnetic field $B_{\parallel}$ on the same 3D grid as our electron density map. We first generate a random Gaussian process which represents the field of $B_{||}$, and convert this to a synthetic Faraday rotation sky map using equation~\ref{eq:rm_forward}. We then adjust the smooth $B_{||}$ field, until its forward modelled RM sky most closely matches the observed RM sky, while keeping the field spatially correlated and respecting the priors listed in table~\ref{priortable}. This is done via geometric variation inference (described below). Our formulation follows recent work that models Galactic 3D fields with Gaussian processes and solves the inference with variational methods (e.g. \citet{edenhofer23,soding25}).

A potential caveat in our approach is that we treat the reconstructed \citetalias{hutschenreuter22} RM sky as if it were direct observational data, using the posterior mean and variance from the IFT analysis as input to our reconstruction. Strictly speaking, we are not using information optimally, since we neglect cross-correlations between RM and dust structure. As well, using inference outputs directly as data likely underestimates the uncertainties due to the implemented methods of approximating the true posteriors. 

However, for the purposes of this work, we expect systematic uncertainties—such as imperfections in the $n_{e}$ map and contamination from outside the reconstructed volume—to dominate over this statistical imprecision. We therefore proceed by adopting the \citetalias{hutschenreuter22} RM sky posterior mean and variance as effective “data” for our analysis.

At an intuitive level, the method works by morphological matching between structures that are seen in both the RM sky and 3D electron density map. When a filament, shell, or HII region is well resolved in the 3D $n_{e}$ map and produces a corresponding feature on the RM map, the optimizer can attribute the RM variation to a local $B_{||}$ in that same volume element. Consequentially, the map is expected to reveal information primarily in Warm Ionized Medium (WIM) structures that are high in $n_{e}$ and distinctive on the sky, and our map confidence is lower elsewhere.

We highlight that this reconstruction probes only the radial component of the GMF, as this dataset does not contain information on the plane-of-sky component of the magnetic field ($B_{\perp}$). While other datasets exist which could help to constrain $B_{\perp}$, incorporating these into a full three-dimensional, three component reconstruction is left for future work.
 
\subsubsection{Priors}

We model $B_{||}$ as a Gaussian random field (GRF) whose spatial correlations are also inferred by the optimiser. For the power spectrum, we use the model as described in \citet{arras22}. This is based off a power law, but allows for fluctuations in the power spectrum function across different size scales. These fluctuations take the form of a smooth bending of the function (controlled by the parameter \emph{flexibility}). Some examples of prior samples of our power spectrum are shown in figure~\ref{fig:powerspec}. We also put priors on the hyperparameters of the amplitude of $B_{||}$ perturbations and mean value of the magnetic field. The amplitude hyperparameter controls the variance of the value of $B_{||}$ throughout the field, and the mean controls the value around which the fluctuations on the GRF occur. These hyperparameters are also inferred by the optimizer while respecting the priors chosen (shown in table~\ref{priortable}).

\begin{table*}
\centering
\caption{Priors for the $B_{||}$ field.}
\begin{tabular}{lcccc}
\hline\hline
Name & Distribution & Mean & Standard deviation & Degrees of Freedom \\
\hline
$B_{||}$ latent grid & Normal & 0 & 1.0 & $256^{3}$ \\
\hline
Mean value of $B_{||}$ ($\mu G$)& Normal & 0.0 & --- & 1\\
Mean $B_{||}$ standard deviation ($\mu G$) & Normal & 5.0 & 3.0 & 1 \\
Amplitude of $B_{||}$ ($\mu G$) & Log-normal & 3.0 & 1.0 & 1 \\
\hline
Power spectrum, mean slope & Normal & $-4.0$ & 0.5 & 1 \\
Power spectrum, flexibility & Log-normal & 0.2 & 0.05 & 1 \\
\hline
\label{priortable}
\end{tabular}
\end{table*}

\subsubsection{Inference with \texttt{NIFTy}/geoVI}

We infer $B_{||}$ by comparing the forward modelled sky (generated via eq.~\ref{eq:rm_forward}) to the RM map and adjusting both the magnetic field and its hyperparameters to improve agreement while respecting the chosen priors. This defines a high-dimensional \emph{posterior} probability distribution (i.e. the probability distribution of the unknown quantities given the data and priors). Direct sampling of this posterior is computationally infeasible due to the high dimensionality of the problem, so we use \emph{geometric variational inference} (geoVI) \citep{frank21} as implemented in \texttt{NIFTY8}. geoVI is a variational inference method for high-dimensional problems that builds on Metric Gaussian Variational Inference (MGVI) \citep{knolmuller19}. MGVI approximates the posterior with a Gaussian distribution whose covariance is informed by the local Fisher information metric. geoVI extends this by introducing a non-linear coordinate transform which changes the shape of the posterior to be better represented by a Gaussian. This coordinate transform is again determined using the Fisher information metric. Further algorithmic details are given in \citet{frank21}.

We use the published per-pixel uncertainties in the \citetalias{hutschenreuter22} RM sky, and propagate uncertainties by drawing posterior samples of $B_{\parallel}$ from the geoVI approximation. We then fold in the uncertainties in the underlying dust map derived from \citet{edenhofer23}. While \citet{mccallum25} models the electron density using the \emph{mean} map of dust density from \citet{edenhofer23}, we first re-run the \citet{mccallum25} radiative transfer code on each of the 12 posterior samples of the \citet{edenhofer23} dust map. This gives 12 distinct $n_{e}$ field reconstructions, one for each dust map sample. For each of these 12 electron-density maps, we run the geoVI optimizer and draw 10 samples from the posterior, for a total of 120 equally likely samples of $B_{||}$. These 120 samples are then used both to report mean and standard-deviation maps of $B_{||}$.

\section{Results}
\label{results}

Figure~\ref{fig:allslices} shows the results of our reconstruction in slices at set values of $z$ above and below the midplane. From the top, the rows are at $z=500$ pc, $250$~pc, $-250$~pc and $-500$~pc. The midplane slice is neglected due to being entirely hidden behind our midplane mask and unconstrained in this inference. Columns show slices through the map of $B_{||}$, the sample-to-sample uncertainty in $B_{||}$, the electron density slice, and the slice of RM contributions obtained as $B_{||}n_{e}$. We also include further slices at $z=100$ pc, $50$~pc, $-50$~pc and $-100$~pc in the appendix. These slices reveal coherent sign structure on large scales, including differences above and below the plane.

In our reconstruction the large-scale pattern of $B_{\parallel}$ is different above and below the plane. Below the plane the magnetic field is dominated by a single, roughly clockwise component. A simple, clockwise magnetic field appears in these figures as a dipole structure, with blue showing the field pointing away when looking in the $+y$ direction, and red showing the magnetic field pointing towards us when looking in the $-y$ direction. Above the plane, however, we see stronger evidence for a change in the sign of $B_{||}$ in the direction of the Galactic centre (manifesting as a quadrupole like structure in the slices above the plane). In addition to the sign changes associated with the quadrupole pattern, we also observe a slight rotation in the structure of $B_{||}$ from that expected from a clockwise magnetic field which is aligned with our Galactic coordinate system. This behaviour is suggestive of a non-zero pitch angle between the large-scale magnetic field and purely circular Galactocentric orbits.

Figure~\ref{fig:reconstructed} shows the mean reconstructed RM sky from the $12\times10$ posterior samples, alongside their standard deviation, and the original RM sky from \citetalias{hutschenreuter22}. Several notable features in the RM map have been successfully recovered, alongside more global trends across the sky. This includes the known symmetries of the quadrupole structure towards the Galactic centre, and dipole towards Galactic anti-centre. We recover an RMS average value for the magnitude of $B_{||}$ throughout the reconstructed volume of $1.63\pm 0.16\rm \mu G$. The RMS value of $B_{||}$ has been evaluated for each of the 120 sample maps, and the reported uncertainty here is the standard deviation of these 120 values.

The relative uncertainty map of the reconstructed sky is dominated by variance in the $n_{e}$ samples, with regions on the edges of ionized volumes showing much higher uncertainties due to the movement of these $n_{e}$ structures from sample to sample. This movement of $n_{e}$ structures is due to the variance in the dust map samples. Because of the nature of the ionization simulations of \citet{mccallum25}, small variations in the underlying \citet{edenhofer23} dust structure can lead to large differences in ionized volumes.

Regions close to field reversals on the sky are also seen to have high relative uncertainties. This is because where the net RM along a sight-line is close to zero, the data mainly indicate that positive and negative contributions cancel, which is consistent with a wide range of local configurations of $B_{||}$ and $n_e$, and therefore leads to a degenerate and poorly constrained $B_{||}$ in those directions.

Figure~\ref{fig:meanbfield} shows a top-down view of $B_{||} n_{e}$ in the local volume. This can be considered a top down version of the \citetalias{hutschenreuter22} sky map, or alternatively as the z-integrated view of the rightmost column in figure~\ref{fig:allslices}. With the region of our reconstruction below the $|b| <5^{\circ}$ mask being unconstrained, we do not include these cells in this plot. The convention of positive RM denoting $B_{||}$ pointing towards the observer is adopted \citepalias{hutschenreuter22}, meaning red on the colour map denotes $B_{||}$ towards the Sun, and blue is away from the Sun. This map is hence showing an integration along the z-axis of the contributions to the \citetalias{hutschenreuter22} RM sky.

Because the magnetic field geometry above and below the mid-plane differs substantially (as seen in Fig.~\ref{fig:allslices}), this full $z$-integration should be interpreted only as a visualization of where the dominant local RM contributions arise. It is not intended to imply vertical symmetry of the magnetic field.

Because we model $B_{||}$ as a correlated Gaussian random field, its statistical structure is fully characterised by its power spectrum $P(k)$. Here $k$ denotes the spatial wavenumber, i.e. the inverse of a physical length scale ($k \sim 1/\ell$), so that small $k$ corresponds to large spatial scales, while large $k$ corresponds to small spatial scales. Physically, the power spectrum therefore quantifies how magnetic field variance is distributed across spatial scales, and its slope encodes whether most of the power resides on large coherent scales or in small-scale turbulent structure. Figure~\ref{fig:powerspec} shows the inferred power spectrum for the field of $B_{||}$. The power spectrum has an average log-log slope of $-2.73 \pm 0.19$ over the full range of k values from $1/9.8~\rm pc^{-1}$ to $1/2.5~\rm kpc^{-1}$. Some fluctuation from a straight power law is preferred by the optimizer, with a slight knee being seen in the posterior samples, at a k-value equivalent to size scales of 80~pc. No sharp features are seen in the inferred power spectrum.

\begin{figure*}
    \centering
    \includegraphics[width=1.0\textwidth]{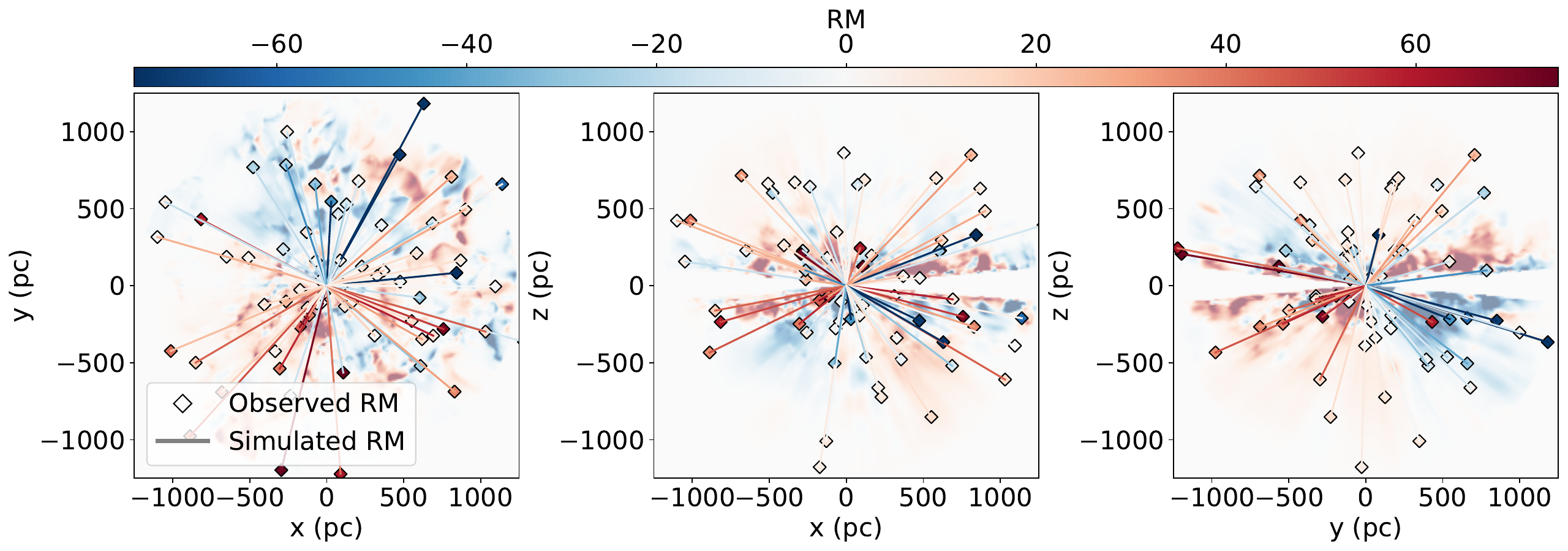}
    \caption{Visualisation of our pulsar validation dataset projected along the 3 principal axes of the reconstructed volume. These pulsars are not included in the \citetalias{hutschenreuter22} reconstruction of the RM sky, or the 3D field of $B_{||}$. Diamonds represent pulsar positions, with their colour representing their observed RM. Lines show the integrated path to each pulsar, with line colours showing the reconstructed RM from our 3D reconstruction. The underlying images shows the z-integrated 3D reconstructed structure of $n_{e}  B_{||}$.}
    \label{fig:pulsars3d}
\end{figure*}

Figure~\ref{fig:rmdm} shows the validation of our results against local Galactic pulsars. The observed and reconstructed pulsar RMs are also shown in figure~\ref{fig:pulsars3d}, in which the positions of the set of pulsars are shown alongside the observed RMs (diamond colours) and the reconstructed RMs (line colours).

The reconstructed RM and DM values show significant scatter, with reduced $\chi^{2}=5.3$ and $5.5$ respectively. We calculate the correlation coefficients for for the RM and DM plots via Monte Carlo sampling from the uncertainties in predicted RM and DM. The find correlation coefficients are $0.79 [0.76,0.82]$ (95\% interval) for RM and $0.73[0.62,0.80]$ (95\% interval) for DM. Because RM depends on both $n_{e}$ and magnetic field, while DM depends only on $n_{e}$, some of the correlation between $RM_{\rm model}$ and $RM_{\rm obs}$ may arise simply from both quantities tracing electron density fluctuations. To quantify how much RM agreement remains after removing this effect, we compute the correlation between $RM_{\rm obs}$ and $RM_{\rm model}$ after first removing the part of each that can be explained by $(DM_{\rm obs},DM_{\rm model})$. This was done by forming the joint $4\times4$ correlation matrix of $(DM_{\rm obs},DM_{\rm model},RM_{\rm obs},RM_{\rm model})$ and isolating the residual RM--RM correlation using the Schur complement of the DM block (i.e.\ removing the DM dependence). The resulting conditional correlation is $\rho_{RM,|,DM}=0.80 [0.76,0.84]$ (95\% interval), similar to the raw RM correlation, indicating that the RM agreement is not solely driven by the DM correlation.

\section{Discussion}
\label{discussion}

Our reconstruction isolates the line-of-sight (LOS) component of the local GMF within 1.25 kpc. The maps show signs of the expected quadrant-by-quadrant coherent structure, but also exhibit signs of a more turbulent, small-scale variation at the 100–500 pc level. The correlation between the forward-modelled RM sky and the input RM map, together with the distance resolved pulsar correlation, indicates that the combination of the dust map based electron density maps and a smooth, correlated $B_{||}$ appear to capture the dominant LOS physics in the local volume.

The above/below-plane asymmetry is consistent with the tilted plane reversal geometry inferred from diffuse-emission and extragalactic RM studies \citep{ordog17,ma20,Booth26}, and with the large-scale longitude-dependent symmetry reported by \citet{dickey22}, who illustrate a $\sin(\ell)$-like RM pattern below the plane and a $\sin(2\ell)$-like pattern above it. Recent pulsar RM work by \citet{curtin24} further suggests that the Local/Sagittarius arm reversal is closer to the Sun at positive latitudes and lies farther away at negative latitudes; within our local box we therefore intersect the reversal primarily for $b>0^{\circ}$, while the region below the plane remains in clockwise-field domain. Our reconstruction thus appears to capture the nearby portion of the same large-scale reversal structure, without imposing it as a parametric prior.

We also see evidence of a non-zero pitch angle between the magnetic field and circular Galactocentric orbits. Such a pitch angle is a generic feature of spiral magnetic geometries and is routinely included in global parametric GMF models (see review of \citet{jaffe19}). Its emergence here, despite the absence of any imposed spiral or parametric field geometry in our prior, indicates that this pitch is driven by the RM data and local electron-density structure. Because our non-parametric framework does not impose a fixed global field geometry, such features are not prescribed a priori by the model parameterization.

As shown by the pulsar validation in Sect.~\ref{results}, the agreement between observed and predicted pulsar RMs remains the same after controlling for the fact that both RM and DM trace fluctuations in the electron density, indicating that the RM agreement is not simply inherited from the electron-density structure. This supports the interpretation that our reconstruction is capturing genuine variations in $B_{||}$ within the local volume. At the same time, the substantial scatter and elevated reduced $\chi^{2}$ values point to residual systematics—most likely from imperfections in the underlying $n_{e}$	model. This is also reflected in the slightly weaker DM-DM correlation relative to the RM--RM comparison. A likely reason is that RM is sensitive to the product of $n_e B_{||}$, and some of the $n_e$ errors are effectively absorbed by compensating shifts in $B_{||}$ during our reconstruction, leading to a tighter apparent RM--RM relation than DM--DM.

Some of the largest sources of systematics in the electron density model are due to the assumed constant dust-to-gas ratio and the incomplete set of ionizing sources. The dust–to–gas conversion in reality should be spatially variable; and having assumed a single ratio will over/under-estimate $n_{e}$ in different environments. Ionization is set by the catalogued O-star population and photoionization only; missing ionizing sources (B stars, sdOBs, hot white dwarfs, and cosmic-ray ionization) will bias $n_{e}$ low in some regions, which in turn biases $B_{||}$ high to match the RM. Conversely, over-estimated $n_{e}$ pushes $B_{||}$ low. Further work on the spatial variability of the cosmic ray ionization rate would also be beneficial in improving these models, as it would give us the ability to include not just O-star photoionized electrons, but also possible contributions from the denser, only partially ionized medium.

Our mid-plane mask reduces contamination from beyond 1.25~kpc, but lines of sight at moderate $|b|$ can still contribute from outside the box; these regions show up as areas of larger uncertainties and residuals in figure~\ref{fig:reconstructed}. In the reconstructed RM sky we also recover a prominent positive (red) RM structure at high latitude associated with the intermediate-velocity H I feature known as the IV arch. This structure was originally identified by \citet{kuntz96}, finding that it lies at a $z$-height of approximately 0.8–1.5~kpc. This would place the IV arch close to the boundary of our reconstructed volume. This structure has since been observed both as a positive RM feature \citep[region~i in Fig.~5 of][]{oppermann12} and in the observed H$\alpha$ sky from WHAM \citep{haffner03}. The IV arch also does not appear in the more locally probing LOFAR Two-metre Sky Survey (LoTSS) data \citep{erceg22}, and therefore represents a plausible example of known contamination from outside the reconstructed volume.

While the IV arch may be an example of outside contamination in our reconstruction, masking regions which are believed to be outside the volume represents a large challenge in itself, requiring a priori knowledge about the distance to every structure in the RM sky. We conservatively mask only the midplane, as we do not believe contamination from distant structures to be a globally dominant systematic error for 2 reasons. Firstly, due to the evidence that much of the RM sky is a local structure \citep{pakmor18,Maconi_2025,korochkin25,pelgrims25}, and secondly because we have recovered valid distance limited information as confirmed by the pulsars in figure~\ref{fig:rmdm}.

We expect that more extended dust maps will be possible in the future, resulting in more extended maps of $n_{e}$ and in turn a more accurate map of local $B_{||}$ with less unknown contamination from outside the reconstructed volume. 

Through isotropy arguments, one can extrapolate our mean RMS strength of $B_{||}$ to approximate a total B-field RMS strength of $2.8\pm 0.3\rm \mu G$. We also calculate a mean magnetic energy density in our reconstruction as $3.1 \times 10^{-13}~ \rm erg~ cm^{-3}$. Our method does not constrain the level of isotropy of the magnetic field, and this result is simply an approximation under this broad assumption.

We additionally note that our results for total magnetic field strength are likely a lower limit, as small-scale, high-amplitude reversals in the magnetic field can cancel out Faraday rotation, leaving these structures undetectable in RM. We rely only on our correlation structure to infill these reversals in a way which is consistent with the RM data. The existence of these small-scale reversals could act as an explanation for much of the scatter in our recovered set of local pulsar RMs.

We constrain only the line-of-sight magnetic field component, $B_{||}$ and not the plane-of-sky component $B_{\perp}$. In principle, future work could involve combining other tracers such as dust or synchrotron polarization, which are sensitive to $B_{\perp}$. While this could provide complementary constraints, these observables likely probe different ISM phases, so combining them requires care.

\subsection{Comparison to other models}
\label{othermodels}

Widely used models fit analytic disk and halo magnetic fields (often with striated and turbulent components) to all-sky extragalactic RMs and polarized  synchrotron emission. These models prove valuable for global features, but by construction exist at kpc resolution and depend strongly on choices for thermal/cosmic ray electron distributions. \citet{planck16} compared a number of models and showed that matching both synchrotron and dust everywhere is difficult and model-dependent. By construction, however, these global comparisons refine only the large-scale picture and do not yield a distance-resolved $B_{\parallel}$ map in the local ISM at parsec scales.

In this section we focus on one of the many available large scale parametric magnetic field models (\citet{unger24}, herein referred to as \citetalias{unger24}). While our conclusions are largely independent of model choice for comparison, we include a further similar set of figures using the models of \citet{korochkin25} in appendix~\ref{korocompare}.

Figure~\ref{fig:bcompare} shows the `base' model of \citet{unger24} (fit to extragalactic RMs and polarized synchrotron intensity maps) in comparison to our reconstruction. By construction, the model of \citetalias{unger24} represents only the large-scale coherent Galactic magnetic field, with local and turbulent contributions effectively averaged out through spatial smoothing and through the use of all-sky, long path-length integrals. In contrast, our reconstruction is explicitly designed to capture local, distance-resolved structure, and therefore naturally recovers substantially higher peak values of $B_{||}$ on small scales. The difference in amplitude between the two models is thus expected and reflects the fundamentally different physical components being probed: a coherent global magnetic field versus a local magnetic field that includes strong turbulent and small-scale contributions.

By the nature of probing smaller scale structures, we measure higher absolute values of $B_{||}$ than \citetalias{unger24}, and for an easier comparison we also include a version of our reconstruction smoothed with a Gaussian kernel of 100~pc width. This smoothing was done in 3D, and the smooth $B_{||}$ map was then integrated along the $z$ axis. We separate this integration into one both above and below the midplane, and also exclude any cells behind our $|b|<5^{\circ}$ midplane mask. In general, the quadrant by quadrant structure predicted by \citetalias{unger24} is recovered in our map, in particular in the positive $z$ region. The structures below the plane show more disagreement, with our reconstruction showing more of a dipole like structure around the Sun, while \citetalias{unger24} maintain a quadrupole on the scales of this volume. 

We also resolve a reversal structure in the upper-right ($0^{\circ} < \ell < 90^{\circ}$) quadrant of the $B_{||}$ field that is not present in the \citetalias{unger24} parametric model (see figure~\ref{fig:bcompare}). We interpret this feature cautiously, as the reversal is not seen as strongly in the corresponding 3D maps of RM contributions. This region contains very low free-electron densities in the $n_{e}$ map, with few ionizing sources in the sample used by \citet{mccallum25}. As discussed above and confirmed through our mock tests (see appendix~\ref{mocktests} for more details), volumes of our reconstruction with very low electron density are only weakly constrained by the data. This possibly unphysical reversal may therefore instead indicate that the true electron density in this region is higher than represented in the current $n_{e}$ model.

Despite the advantages in our reconstruction of higher resolution and limited parametric assumptions, we are still susceptible to errors in our underlying model of $n_{e}$ from \citet{mccallum25}. There are known issues with the underlying model, but future datasets and improved modelling will continue to constrain our 3D electron density maps, and thus our models of the $B_{||}$. There is also scope in the future for joint reconstruction of electron density with $B_{||}$, leveraging datasets probing not only RM, but DM and emission measure from free-free emission, and Faraday-rotated diffuse polarized synchrotron emission.

A comparable approach to our work is presented in \citep{hutschenreuter24}. This work attempts to disentangle the Faraday rotation sky of \citepalias{hutschenreuter22} into a DM and LoS-averaged magnetic field component on the sky, using additional information on the electron density from pulsar DMs and collisional processes such as H$\alpha$ and free-free emission. They find an all sky RMS value of 1.1 $\mu$G for the LoS-average of the radial magnetic field component, with a maximum of 5.4 $\mu$G, which is comparable to our work, with an RMS $B_{||}$ of 1.63 $\mu$G. The electron density information in \citet{hutschenreuter24} is fully complementary to our approach, but many other components of the analysis are very similar. We hence attribute mismatches to the possibly missing electrons in \citet{mccallum25}, or alternatively biases in the electron modelling of \citet{hutschenreuter24}.

\begin{figure*}
    \centering
    \includegraphics[width=1.0\textwidth]{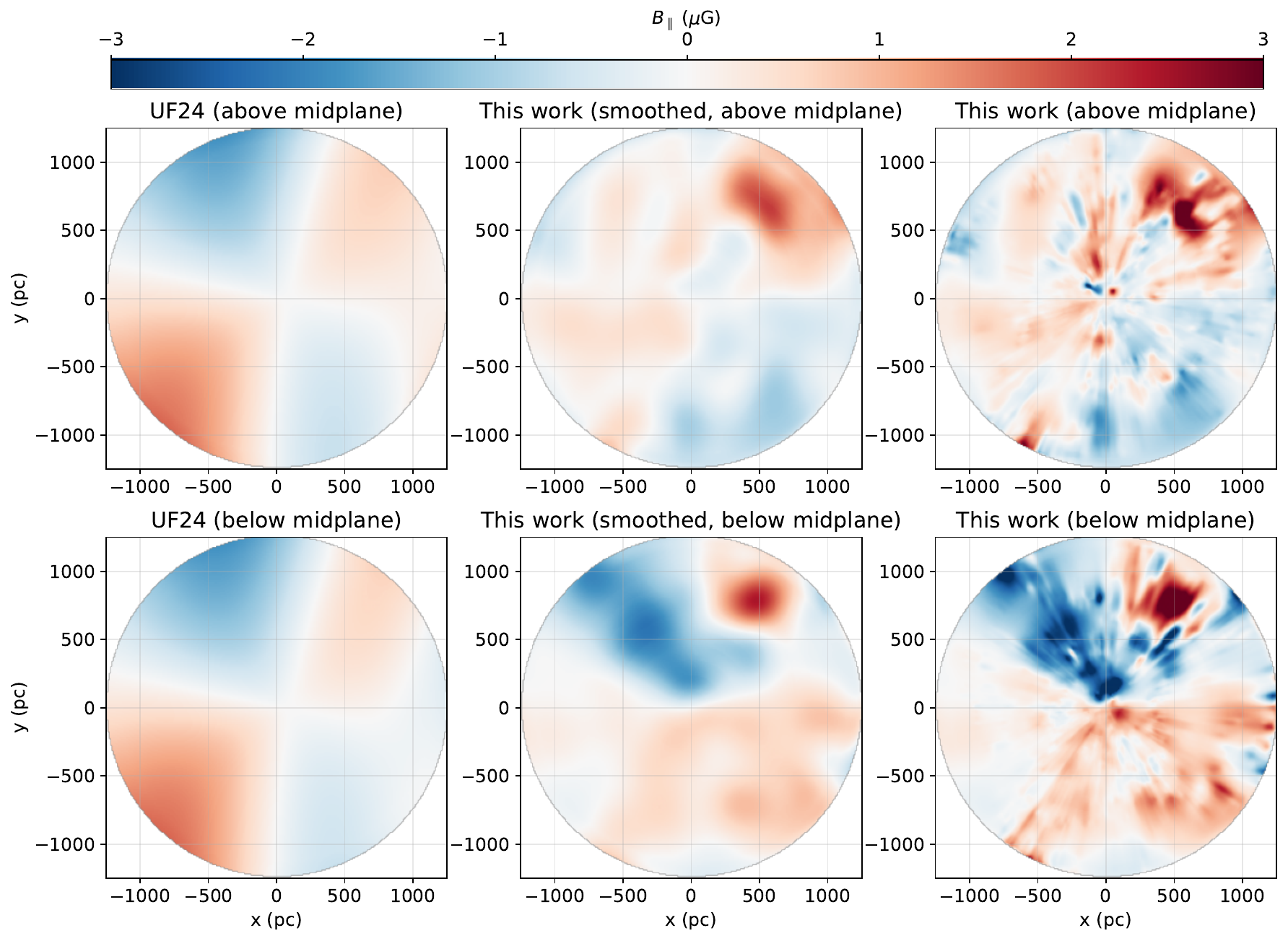}
    \caption{Comparison of the face-on volume weighted mean field of $B_{||}$ from \citetalias{unger24} (left panels), as well as our reconstruction both smoothed with a Gaussian kernel of width 100~pc as described in section~\ref{othermodels} (centre column), and at full resolution (right panel). In all panels, we show the volume weighted mean magnetic field through the z-axis, with the Galactic centre lying on the y-axis to the right. The top row shows the $B_{||}$ field averaged above the midplane to a height of $z=500~\rm pc$, and the bottom panels show the average from the midplane to $z = -500~\rm pc$.  In all panels, only cells above our $|b| < 5^{\circ}$ mask are included in the average.}
    \label{fig:bcompare}
\end{figure*}

\begin{figure*}
    \centering
    \includegraphics[width=1.0\textwidth]{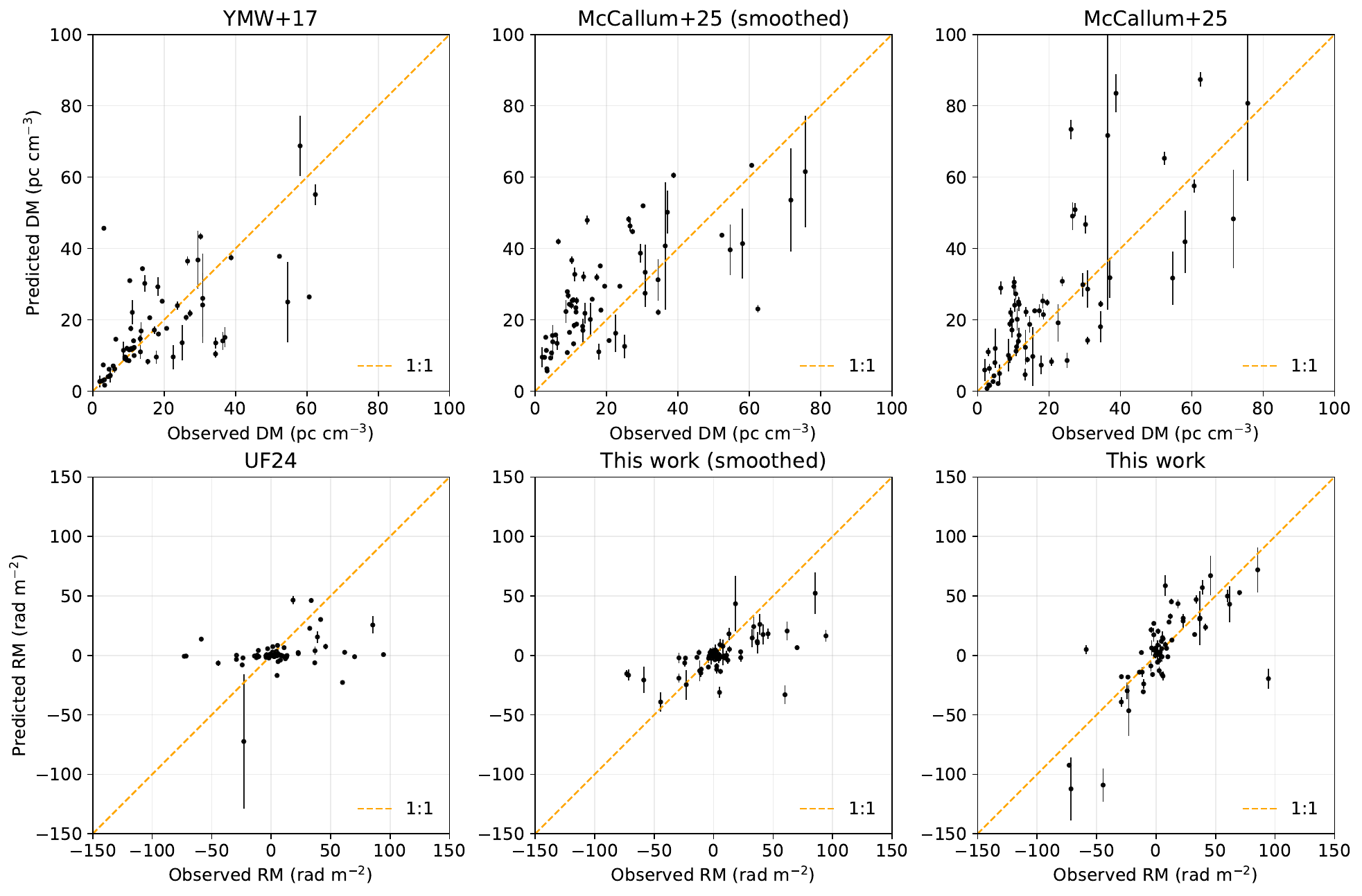}
    \caption{Comparison of DMs predicted by \citet{mccallum25} versus the \citet{yao17} (YMW+17) electron density model used by \citetalias{unger24} (top row), and comparison of RMs predicted by the \citetalias{unger24} `base' model magnetic field with our reconstructed $B_{||}$ field (bottom row). RMs in the \citetalias{unger24} panel have been evaluated with the \citet{mccallum25} electron model. Left panels show the large scale parametric models of \citet{yao17} and \citetalias{unger24}, the middle panels show this work (as well as \citet{mccallum25}) smoothed with a Gaussian kernel of width 100pc, and right panels show this work (and \citet{mccallum25}) with no smoothing. These comparisons show that without the high amplitude small-scale structure of the B-field, the local pulsar RMs cannot be reconstructed. It is also seen that scatter in DM plots is mostly independent of model choice.}
    \label{fig:unger_rms}
\end{figure*}

\subsection{Reliability of small-scale (<100~pc), and radially resolved structure}

One of the novel features of this method and the resulting map is its ability to resolve the Galactic magnetic field at smaller scales than previously possible, due to detail from the highly-resolved map of electron density. In order to test whether the structures we have recovered are data-driven or prior-driven, we have presented a series of validating tests for both the small-scale structure, and the radially-resolved structures. In order to test if our found correlation between observed and predicted RMs to local pulsars (as seen in figure~\ref{fig:rmdm}) is driven by our resolved small scale structure, or simply the coherent magnetic field, we produce figure~\ref{fig:unger_rms}. This figure shows the correlation recovered using the coherent magnetic field model of \citetalias{unger24}, and the electron density map from \citet{mccallum25}. We find that the smooth, large scale magnetic field of \citetalias{unger24} is approximately an order of magnitude too small to explain the local structure in RM as probed by these local pulsars. In combination with the correlation between observed and reconstructed DMs from the same pulsars (or at minimum, the match in order of magnitude in DM), this suggests that the local structure of RM is dominated by small-scale, high-amplitude variations in the magnetic field which are not recovered in these large scale parametric models. This result is not an invalidation of the \citetalias{unger24} model, but rather evidence for valid small scale structure in our recovered B-field. The models of \citetalias{unger24} do not seek to model this turbulent component, and we equivalently show in figure~\ref{fig:unger_rms} that when our local field of $B_{||}$ is smoothed to a similar length scales, we also have too weak a B-field to explain the local pulsar RMs. Also shown in figure~\ref{fig:unger_rms}, for this set of local pulsars, the trends in observed versus predicted DM are seen to be mostly independent of model choice (\citet{mccallum25} versus the \citet{yao17} model employed by \citetalias{unger24}), or level of smoothing. This further suggests that the RM structure we recover is driven by fluctuations in in the magnetic field rather than $n_{e}$. While the \citet{yao17} electron density model produces a slightly cleaner match to the observed pulsar DMs than \citet{mccallum25}, many of these DMs were included in the fitting of the \citet{yao17} model, whereas the \citet{mccallum25} model is constructed independently of the DM data.

It can also be seen in figure~\ref{fig:powerspec} that our reconstruction results in a power spectrum with a mean slope of $-2.73 \pm 0.19$, this is a much shallower spectrum than our prior, suggesting that these small scale fluctuations are likely data driven rather than prior driven. To test the sensitivity of this result on our power spectrum prior, we carried out an identical set of reconstructions using a prior slope of -2.0 instead of our fiducial value of -4.0 (see table~\ref{priortable}). We find no meaningful difference in the reconstructed power spectrum under this change, suggesting we are largely insensitive to our power spectrum prior choice.

Our reconstructed power spectrum slope of $-2.73 \pm 0.19$, while shallower than the spectrum which is derived from \citet{kolmogrov41} turbulence ($-11/3$), is not inconsistent with other observational efforts. \citet{han04} used pulsar RM and DMs to observationally constrain a 1D power spectrum slope over a spatial scale of 0.5 to 15~kpc, finding a slope of -0.37, equivalent to a slope of -2.37 in 3D. They also report evidence for a discontinuity in the power spectrum slope somewhere between a spatial scale of 80~pc and 500~pc. These results are consistent with our reconstructed power spectrum, and the slight steepening of the spectrum we resolve at a spatial scale of approximately 80~pc. In theory our power spectrum slope is a measure of the full 3D magnetic field power spectrum, as adding more components of the B-field only changes the normalization of the spectrum. This extrapolation however relies on the assumption that the field is isotropically oriented amongst the 3 vector components, an assumption likely to break down given the large scale anisotropy of the GMF.

As further validation of the method beyond the mock test, the correlation of observed to reconstructed pulsar RMs, and the lack of RM correlation from the underlying coherent component of the $B_{||}$ field, we also include a test to ensure we are resolving the local pulsar RM structure beyond what is already encoded in the underlying \citetalias{hutschenreuter22} map. To do this, we evaluate the goodness of fit to the pulsar data as the sum of the residual error:

\begin{equation}
{\rm RE}
= \sum_{\text{pulsars}}
\left\lvert 
\frac{ RM_{\rm obs} - RM_{\rm sim} }{ RM_{\rm obs} }
\right\rvert 
\end{equation}

\begin{figure}
    \centering
    \includegraphics[width=1.0\columnwidth]{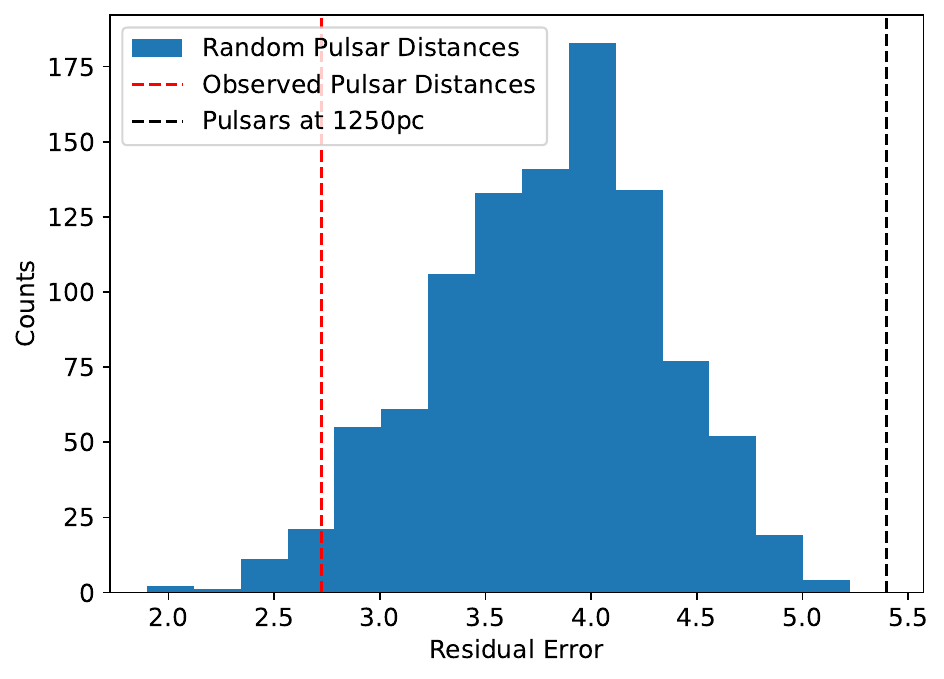}
    \caption{Histogram of residual errors from our pulsar distance test. The leftmost dashed line shows the residual error for our full pulsar RM predictions, and the rightmost line shows the residual error taking RMs as the fully integrated reconstructed map to 1250~pc.}
    \label{fig:histogram}
\end{figure}

We then evaluate the goodness-of-fit for the case of 1) using the observed pulsar distances 2) placing every pulsar at a distance of 1.25~kpc (near equivalent to extracting RM values directly from the \citetalias{hutschenreuter22} map) and 3) use a series of random pulsar distances, sampled uniformly in distance between 69 and 1250~pc (the radial limits of our reconstruction). We carry out 1000 random trials of case 3), and the histogram of the residual errors is shown in figure~\ref{fig:histogram}. We find that using the correct pulsar distances gives us a better fit than 96.8\% of the 1000 trials, and that evaluating pulsar RMs as the full integral to the edge of the reconstructed volume gives a worse fit than all 1000 trials.

\subsection{Future work with diffuse emission Faraday rotation}

In this work $B_{||}$ is constrained using extragalactic source RMs, which provide information about the net Faraday rotation along each LOS. Different $B_{||}$ LOS profiles, with reversals or small-scale fluctuations in magnetic field strength, can produce degenerate RM values. One possible method for breaking these degeneracies is to incorporate diffuse Galactic polarized emission into the forward modelling. Synchrotron emission is ubiquitous in the Galaxy, and its polarization is also subject to Faraday rotation, resulting in complicated profiles of mixed emission and rotation along any LOS. The broadband, single-antenna polarization data sets from the Global Magneto-Ionic Medium Survey \citep[GMIMS;][]{Wolleben19,woll21,Sun25,Ordog26} provide complementary information to the extragalactic RMs through Faraday synthesis \citep[also known as RM synthesis;][]{brentjens05}. Faraday synthesis yields spectra of observed polarized emission as a function of Faraday depth, allowing for sensitivity to multiple Faraday rotation components along the LOS, which are subject to averaging when probed by extragalactic RM sources. These Faraday depth spectra can provide additional constraints on $B_{||}$ by requiring the spectra recovered from the model to match the observations. Complications arise due to frequency- and instrument-dependent depolarization effects that limit the depths that diffuse emission polarization surveys probe. This \textit{polarization horizon} \citep{Uyaniker03} has been estimated by \cite{dickey19} for two of the GMIMS datasets using spatial variability in the first moments of Faraday depths and comparisons with pulsar RMs. Even with good polarization horizon estimates, mapping between Faraday depth and physical depth is usually not straightforward. However, this can at least partially be overcome by correlating the two-dimensional morphologies of Faraday rotation features with structures in other tracers at known distances, for example the GMIMS Faraday rotation and \citet{edenhofer23} dust map comparison in \cite{Booth26}. Further constraining $B_{||}$ using diffuse emission Faraday spectra is beyond the scope of this work, but will be incorporated in future models, which we expect will produce improvements in mapping $B_{||}$ variations and reversals along the LOS.

\section{Conclusions}
\label{conclusions}

By combining a 3D dust map informed $n_{e}$ map with the Faraday rotation sky in a forward model, we recover the radial component of the Galactic magnetic field, $B_{||}$, across the local (within 1.25~kpc) region of the Milky Way. The \citetalias{hutschenreuter22} RM sky reconstruction matches the input map while producing uncertainty that naturally tracks where the data are informative (mid-latitudes, higher $n_{e}$).

Local pulsar RMs, not used in the fit, are predicted by integrating $n_{e}B_{||}$ to the pulsar distances, and are seen to correlate well with observations, albeit with some significant scatter. Predicted DMs from the underlying $n_{e}$ map also correlate with measured DMs. These checks indicate the predictive power of our model, and the conditional covariance coefficient of 0.75 shows that the RM agreement is not an artifact of the $n_{e}$ correlation alone. 

The dominant systematic is the underlying electron-density model from \citet{mccallum25}. Likely biases in this model are the uniform dust–to–gas ratio and a limited ionizing source census (O stars only). Underestimating $n_{e}$ on any one sight-line pushes the optimized $B_{||}$ high, and vice versa. Additional ionization channels (B/sdOB stars, hot WDs, cosmic rays) are missing in places, with uncertain weights of contribution. Our $|b| < 5^{\circ}$ mask reduces contamination from beyond the 1.25~kpc box, but some low-latitude sight-lines still accumulate non-local contributions, reflected in larger uncertainties near the plane.

We regard this model as a first step in more highly resolved mapping of the local Galactic magnetic field. This work is crucially dependent on the underlying models of electron density, which are in turn dependent on the high-resolution dust maps, and ionizing source catalogues/models. With \textit{Gaia} DR4 on the horizon, and recent advancements in OB star catalogues \citep{pantaleoni25}, this pipeline is set to improve from the bottom up. We expect uncertainties to decrease, and map coverage to expand in the near future as a function of data availability. Further advances in RM data from surveys such as POSSUM \citep{gaensler25} will deliver an increase in the surface density of extragalactic RMs by more than an order of magnitude. This improvement in RM-grid sampling will be particularly powerful for our method, as it directly enhances sensitivity to small-scale angular structure in the RM sky and therefore tightens constraints on the local, turbulent component of $B_{||}$. Early surveys such as SPICE-RACS already demonstrate how increased RM sampling density begins to reveal fine angular-scale RM structure that is currently unresolved in existing all-sky reconstructions. Beyond this, advancements in other forms of 3D mapping, such as gas emission mapping \citep{soding25}, will also help to constrain uncertainties on local measurements of the the dust-to-gas ratio, as well as expanding our view of the interstellar density field well beyond the current 1.25~kpc region explored in this work. These advances will in turn improve our $n_{e}$ maps, and thus our understanding of the local Galactic magnetic field.

\section*{Acknowledgements}

This work benefited from the conference ``Structure and polarization in the interstellar medium: A Conference in Honor of Prof. John Dickey," a hybrid meeting hosted jointly at Stanford University and at the Australia Telescope National Facility in February 2025. We acknowledge support from the National Science Foundation (NSF award No. 2502957), from the Kavli Institute for Particle Astrophysics and Cosmology, from the Commonwealth Scientific and Industrial Research Organization, and from the Australian Research Council.

A.K.S. acknowledges support for this work was provided by NASA through the NASA Hubble Fellowship grant HST-HF2-51564.001-A awarded by the Space Telescope Science Institute, which is operated by the Association of Universities for Research in Astronomy, Inc., for NASA, under contract NAS5-26555. A.S.H. acknowledges the support of a Canadian National Scientific and Engineering Research Council Discovery Grant. A.O. was partly supported by the Dunlap Institute at the University of Toronto. The Dunlap Institute is funded through an endowment established by the David Dunlap family and the University of Toronto.

Co-funded by the European Union (ERC, ISM-FLOW, 101055318). 
Views and opinions expressed are, however, those of the author(s) only and do not necessarily reflect those of the European Union or the European Research Council. 
Neither the European Union nor the granting authority can be held responsible for them. 

The authors acknowledge Interstellar Institute's program "ii7" and the Paris-Saclay University's Institut Pascal for hosting discussions that nourished the development of the ideas behind this work.

\section*{Data Availability}

Our 3D map of mean and standard deviation of $B_{||}$ is available at \url{https://doi.org/10.5281/zenodo.19370313}. An accompanying electron density map is also shipped for the purposes of identifying regions which are actually constrained by our method.



\bibliographystyle{mnras}
\bibliography{biblio} 




\appendix

\section{Mock Data - Proof of Concept}
\label{mocktests}

In order to explore the applicability of this method for reconstructing the $B_{||}$ field from the RM sky, we set up a simple test in the form of a mock-up. We generate a Gaussian random field which we attempt to reconstruct. We first pass the random field (mock $B_{||}$ ground truth) through the forward model to give an RM sky, add noise to the generated RM sky, and use this as the data product from which the $B_{||}$ field is recovered. The noise sky was determined from the posterior samples of the \citetalias{hutschenreuter22} by taking a single realization of the RM sky, and then subtracting the mean RM sky. This gives a single sample of a noise sky which adheres to the correlations genuinely detected in the RM across the sky.

To account for the variance in the assumed $n_{e}$ map in our mock test, we generate the RM sky using one realization of $n_{e}$, and then reconstruct using a different $n_{e}$ realization. This should introduce variance in the underlying $n_{e}$ assumption which is consistent with our formalized understanding of the uncertainties in $n_{e}$. 

\begin{figure*}
    \centering
    \includegraphics[width=1.0\textwidth]{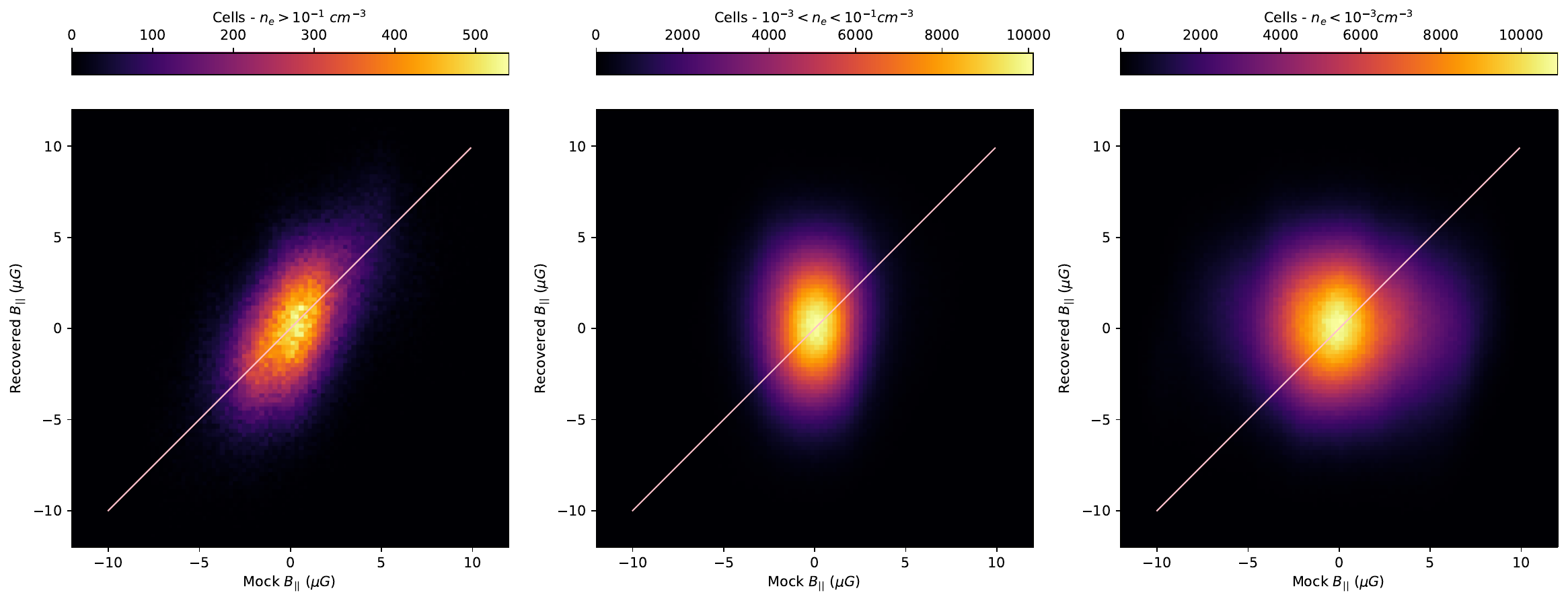}
    \caption{Results of our mock data test. Each panel shows 2D histograms showing the match between the ground truth map, and the map reconstructed from the synthetic RM sky. Panels are separated by underlying electron density in those voxels. The left panel shows high electron density ($n_{e} > 10^{-1} \rm ~cm^{-3}$), middle panel shows medium electron density ($10^{-3} {\rm ~cm^{-3}} < n_{e} < 10^{-1} \rm ~ cm^{-3}$), and the right panel shows only voxels of low electron density ($n_{e} < 10^{-3} \rm ~cm^{-3}$). The match shows a strong correlation for high electron densities, but is poor for medium densities, and very poor for low electron densities. This aligns with expectation, as these low density volumes are unconstrained by our method.}
    \label{fig:mockscatter}
\end{figure*}

Figure~\ref{fig:mockscatter} shows the results of this test in the form of 2D histograms showing per voxel the reconstructed value of $B_{||}$ versus the underlying $B_{||}$ which generated the mock RM sky. We separate the voxels into three histograms by their $n_{e}$ value. We find that the reconstruction works well at higher electron densities, but where there are fewer free electrons in the $n_{e}$ model, the structure of $B_{||}$ is less constrained by the data. This is to be expected, and is a fundamental limitation of using Faraday rotation to measure the magnetic field. When $n_{e}$ is close to zero, the RM contribution from those voxels is also very small and contains less information about the magnetic field. We remain aware of this insight throughout our results section, discussion and conclusions.

\begin{figure*}
    \centering
    \includegraphics[width=1.0\textwidth]{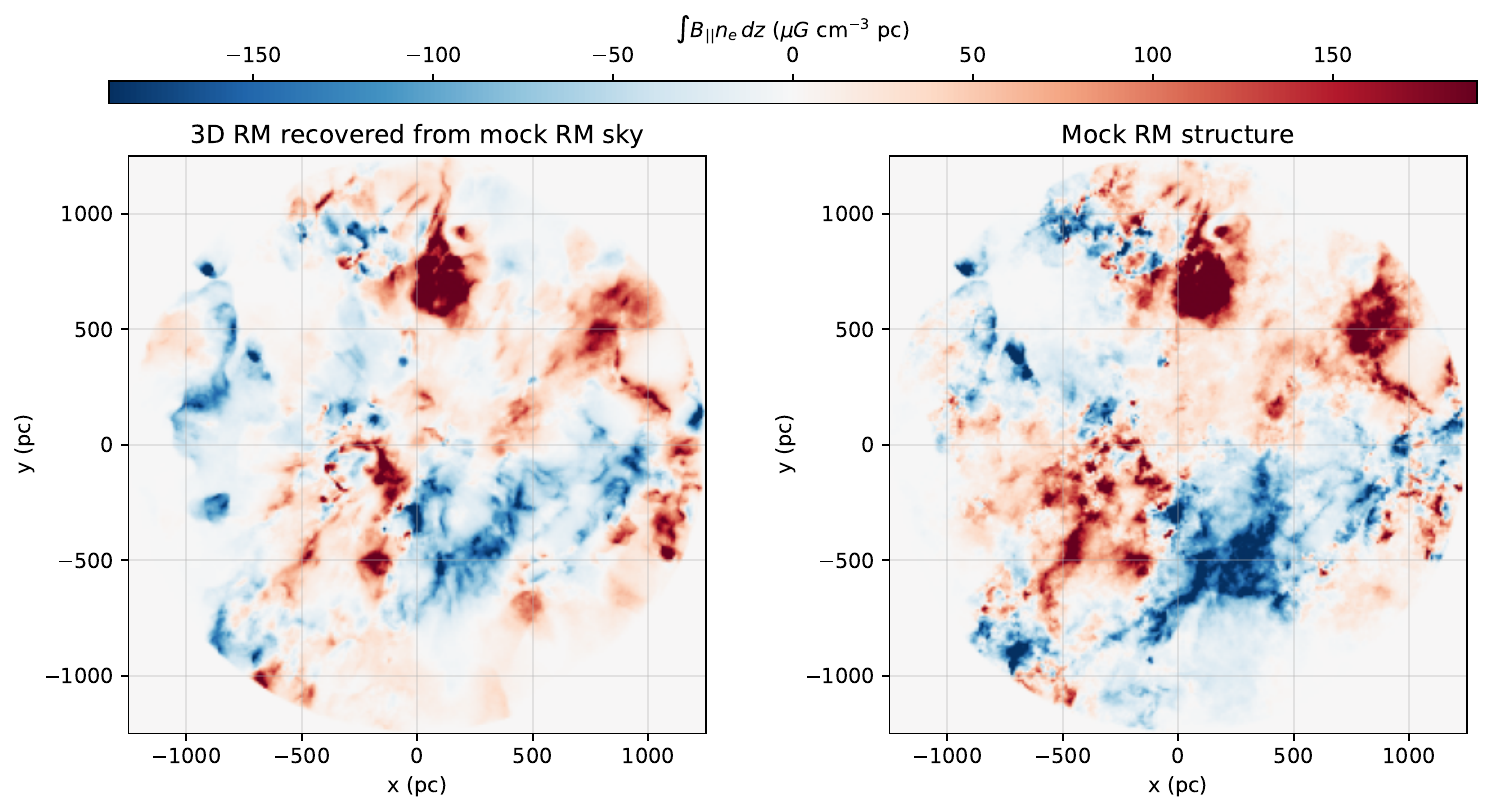}
    \caption{The results of our mock test. Right panel is the underlying RM structure from which the mock data was generated. The left panel shows the result of our reconstruction in 3D. These plots are equivalent in structure to figure~\ref{fig:meanbfield}, whereby we have integrated the differential RM contributions through the z-axis, while exuding any voxels behind the $|b| < 5^{\circ}$ midplane mask.}
    \label{fig:mockcompare}
\end{figure*}

Figure~\ref{fig:mockcompare} shows the top-down view of the mock and reconstructed RM structure from this test. The two structures are seen to match very well, confirming that with an accurate enough map of $n_{e}$, and small enough contamination from outside the reconstructed volume, our technique is methodologically sound.

We also have carried out a mock test where the electron density structure was assumed to be known to 100\% accuracy. This test results in a near perfect match to the ground truth above electron densities of $10^{-2} \rm ~cm^{-3}$. This further confirms that our method works well for these densities, and that for much lower electron densities, the RM dataset is not constraining the magnetic field.


\section{Other Reconstructed Slices}

\begin{figure*}
    \centering
    \includegraphics[width=1.0\textwidth]{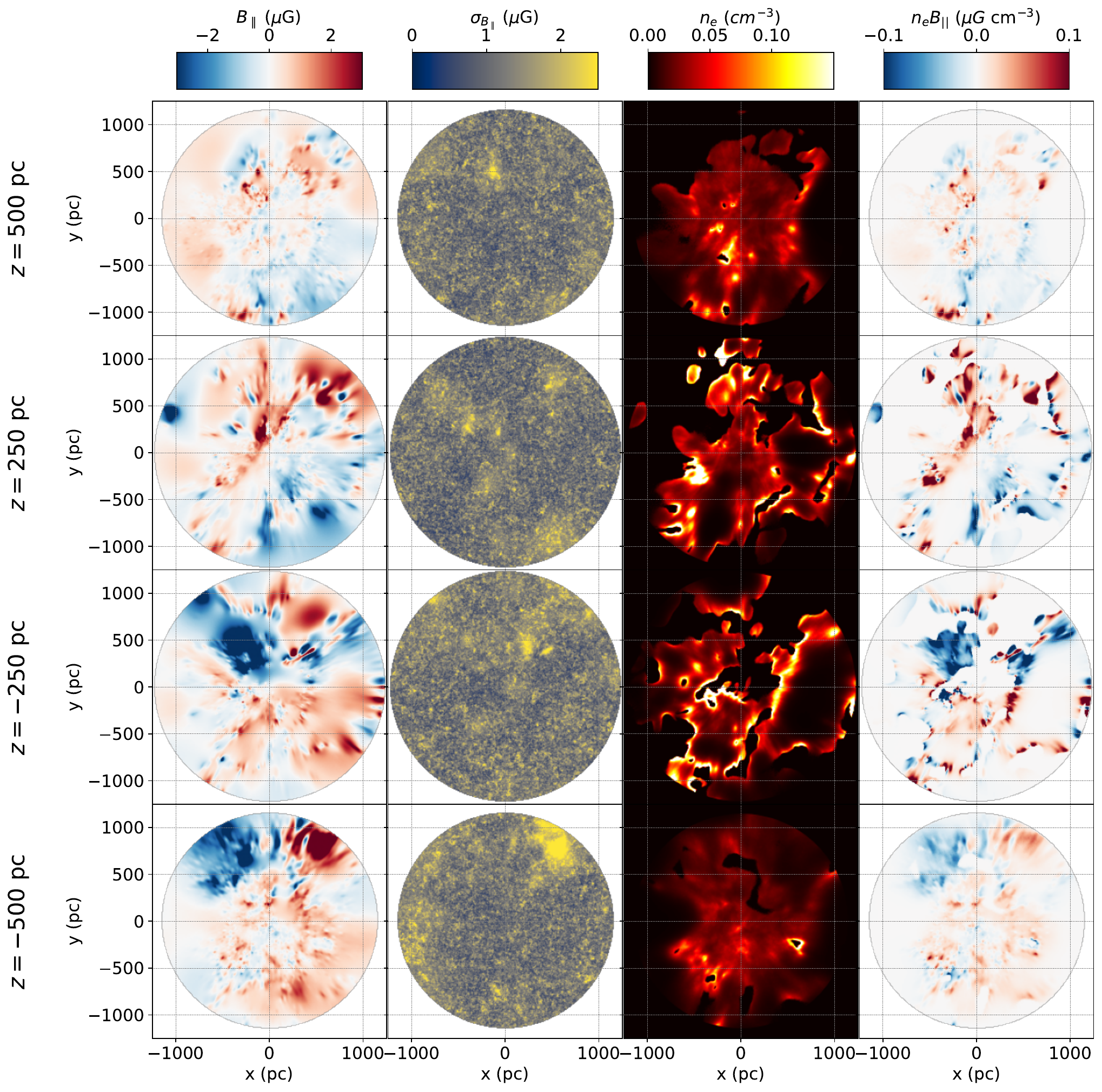}
    \caption{The same figure as fig~\ref{fig:allslices}, but at $z$-values of $100$~pc, $50$~pc, $-50$~pc and $-100$~pc. The colour scales are the same apart from the left column ($B_{||}$), which covers a larger range of $B_{||}$ values.}
    \label{fig:lowal_slices}
\end{figure*}

In figure~\ref{fig:lowal_slices} we include further slices at $z$-values of $100$~pc, $50$~pc, $-50$~pc and $-100$~pc. We exclude any cells behind our $|b| < 5^{\circ}$ midplane mask, meaning closer slices to the midplane show a smaller reconstructed area.

\section{Comparison to Korochkin et al. (2025)}

\label{korocompare}

Here we reproduce Figures \ref{fig:bcompare} and \ref{fig:unger_rms} for comparison with the global parametric model of \citet{korochkin25}.

\begin{figure*}
    \centering
    \includegraphics[width=1.0\textwidth]{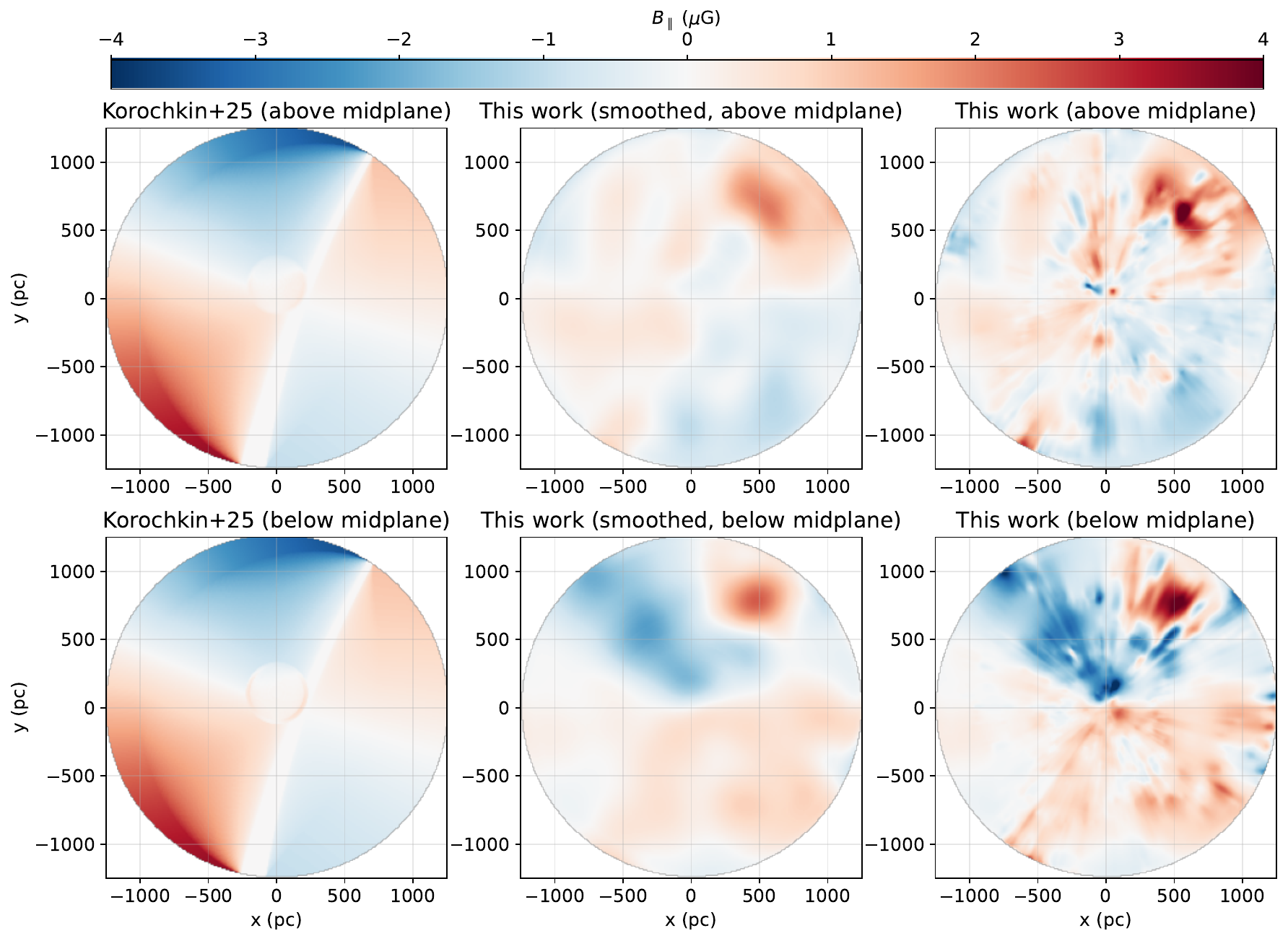}
    \caption{Comparison of the face-on volume weighted mean field of $B_{||}$ from \citet{korochkin25} (left panels), as well as our reconstruction both smoothed with a Gaussian kernel of width 100~pc as described in section~\ref{othermodels} (centre column), and at full resolution (right panel). In all panels, we show the volume weighted mean of the $B_{||}$ field through the z-axis, with the Galactic centre lying on the y-axis to the right. The top row shows the $B_{||}$ field averaged above the midplane to a height of $z=500~\rm pc$, and the bottom panels show the average from the midplane to $z = -500~\rm pc$.  In all panels, only cells above our $|b| < 5^{\circ}$ mask are included in the average.}
    \label{fig:bcompare_koro}
\end{figure*}

\begin{figure*}
    \centering
    \includegraphics[width=1.0\textwidth]{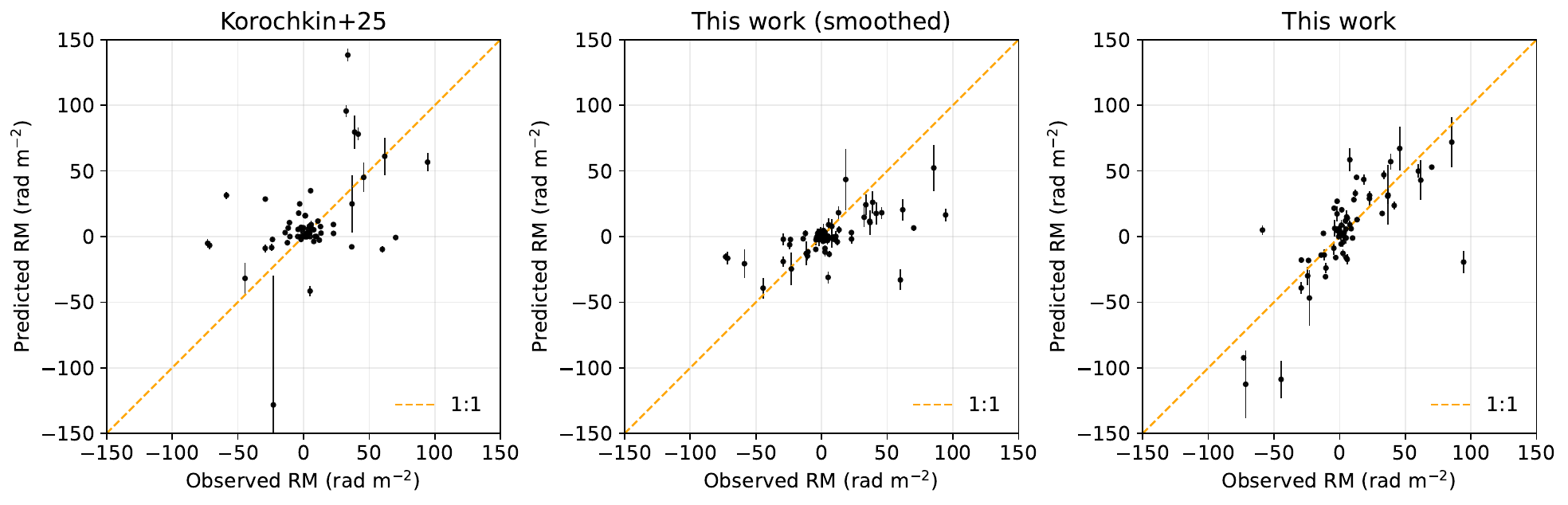}
    \caption{Comparison observed RMs to reconstructed RMs in the \citet{korochkin25} B-field model (left panel), our $B_{||}$ reconstruction smoothed with a 100~pc wide Gaussian kernel as described in section~\ref{othermodels} (middle panel), and our reconstructed $B_{||}$ at full resolution. These comparisons show that without the high amplitude small-scale structure of the B-field, the local pulsar RMs cannot be reconstructed.}
    \label{fig:koro_rms}
\end{figure*}

\bsp	
\label{lastpage}
\end{document}